\documentstyle[prl,aps]{revtex}

\def\be{\begin{equation}}
\def\ee{\end{equation}}
\def\bea{\begin{eqnarray}}
\def\eea{\end{eqnarray}}

\def\ignore#1{ }

\begin{document}
\author{L.P. Yatsenko \thanks{%
Permanent address : Institute of Physics, Ukrainian Academy of Sciences,
prospekt Nauki 46, Kiev-22, 252650, Ukraine}, T. Halfmann, B.W. Shore 
\thanks{%
Permanent address: Lawrence Livermore National Laboratory, Livermore, CA
94550, USA } and K. Bergmann}
\address{{\em Fachbereich Physik der Universit\"at, 67653 Kaiserslautern, }}
\title{Photoionization Suppression by Continuum Coherence: \\
Experiment and Theory }
\maketitle

\begin{abstract}
We present experimental and theoretical results of a detailed study of
laser-induced continuum structures (LICS) in the photoionization continuum
of helium out of the metastable state \mbox{2s ${}^1$S$_0$}. The continuum
dressing with a 1064 nm laser, couples the same region of the continuum to
the {4s ${}^{1}$S$_{0}$} state. The experimental data, presented for a range
of intensities, show pronounced ionization suppression (by as much as 70\%
with respect to the far-from-resonance value) as well as enhancement, in a
Beutler-Fano resonance profile. This ionization suppression is a clear
indication of population trapping mediated by coupling to a contiuum. We
present experimental results demonstrating the effect of pulse delay upon
the LICS, and for the behavior of LICS for both weak and strong probe
pulses. Simulations based upon numerical solution of the Schr\"{o}dinger
equation model the experimental results. The atomic parameters (Rabi
frequencies and Stark shifts) are calculated using a simple model-potential
method for the computation of the needed wavefunctions. The simulations of
the LICS profiles are in excellent agreement with experiment. We also
present an analytic formulation of pulsed LICS. We show that in the case of
a probe pulse shorter than the dressing one the LICS profile is the
convolution of the power spectra of the probe pulse with the usual Fano
profile of stationary LICS. We discuss some consequences of deviation from
steady-state theory.

\noindent PACS number(s): 42.50.Hz , 32.80.Fb, 32.80.Qk
\end{abstract}


\newpage

\newpage

\section{Introduction}

\subsection{Continuum coherence}

The properties of the continuum of quantum states observed in the
photoionization of an atom or the dissociation of a molecule have attracted
interest since the first formulations of quantum theory. Until the
widespread use of lasers, the continuum was regarded as a dissipative
system, to be treated by means of rate equations (typically using the
approximation of the Fermi Golden Rule \cite{Mes61} to calculate the
transition rate into the continuum). Contemporary views no longer regard
continuum states as an irreversible drain of probability. An electron acted
upon by a laser field of sufficiently high frequency to produce ionization
does not move irreversibly away from the binding region \cite
{Sch90,Kul91,Kra92}. Laser interactions with a continuum have been shown to
exhibit Rabi-type oscillations (if the laser field is sufficiently strong) 
\cite{Gol78b,Fri96} 
and to enable nearly complete population transfer into a continuum (i.e.
photoionization or photodissociation \cite{Var96}) 
or from a continuum (i.e. photorecombination \cite{Var97}). 
Both theoretical \cite{Var96,Den85a,Pas97,Pas98,Yat97} 
and experimental \cite{Fri96,Fau93,Fau93a} 
%
work has shown that the wavefunction of an electron in a laser field may
often maintain coherence during the course of photoionization, and that
photodissociation can also maintain coherence. Consequences of continuum
coherence are to be found in the well-studied phenomena of autoionization 
\cite{Fan61,Fan65,Sho67}, 
which is revealed as a resonance in plots of photoionization cross-sections
versus frequency. Such resonances display the effect of interference between
two ionization channels 
whose destructive interference can lead to complete suppression of
photoionization at a specific wavelength.

One of the more interesting coherent phenomena involving the continuum is
the use of a laser field to embed an autoionization resonance into an
otherwise featureless photoionization continuum. This dressing of a
continuum by a bound state by means of a laser field is termed a ``laser
induced continuum structure" (LICS) \cite
{Hut88,Tan89,Sha91,Cav91,Cav95b,Cav95a,Era97,Hal98}.

As with other types of resonances described by scattering theory, LICS can
be observed in various reaction channels. In the LICS observed in
photoionization a strong laser field, connecting an unpopulated bound state
to the continuum, mixes some bound-state character into the continuum; it
places a resonance into the continuum. This optically induced resonance can
be detected by a second (possibly weaker) laser field which serves as a
probe and induces a transition from a populated bound state into the newly
structured continuum.

We remark that when the continuum is composed of several channels (derived
e.g. from different asymptotic internal states of dissociation products) the
dissociation has been shown to give rise to a quantum control scheme \cite
{Che95b} 
in which the ratio between the asymptotic channels can be controlled by
tuning the dissociating (or dressing) laser across a LICS resonance.

Theoretical descriptions of continuum structure near autoionizing resonances
have been presented \cite{Lam81,Pav81,Rza81,Rza83,Lam89} as have LICS in the
absence of such resonances \cite{Lam86a,Lam86b}. Observations of LICS in
photoionization near autoionization resonances \cite
{Fau93,Fau93a,Fau94,Kar95} 
as well as observations of LICS in photoionization to an otherwise {\em %
unstructured} continuum \cite
{Hut88,Tan89,Sha91,Cav91,Cav95b,Cav95a,Era97,Hal98} 
have been reported recently. Under appropriate conditions this LICS can lead
to the suppression of photoionization, by creating a coherent-superposition
of bound states immune to photoionization (a trapped state, or dark state 
\cite{Ari96}). 
The occurrence of significant photoionization suppression signals the
creation of a dark state; only with coherent excitation can this aspect of
continuum coherence be revealed.

\subsection{The resonance profile}

When a coherent radiation field resonantly excites a two-state system, then
a probing of either energy level by a weak probe field reveals a splitting
(the Autler-Townes splitting) attributable to the dressing of bare atomic
states by a strong field \cite{Sho90}. 
When the strong transition connects a bound (but unpopulated) state to a
continuum, a weak probe will again reveal structure, the LICS.

The theoretical basis for describing LICS has been presented in numerous
articles (see for example \cite{Kni84,Kni90}), 
for cases in which the fields are constant-amplitude continuous-wave
monochromatic, and for pulsed radiation. As such work has shown, the LICS of
photoionization typically takes the form of an asymmetric window resonance,
for which the frequency dependence of the photoionization cross-section $%
\sigma(\omega)$ may be parameterized by the form of a typical scattering
resonance in a background \cite{Sho67,Sho68} 
\begin{equation}  \label{BWprofile}
\sigma(\omega) = C(\omega) + \frac{ B + A x} {x^2 + 1}
\end{equation}
where $x \equiv (\omega - \omega_R)/(\Gamma/2)$ is a dimensionless detuning
of the light frequency $\omega$ away from the resonance value $\omega_R$,
measured in units of the resonance-width parameter (or loss rate) $\Gamma$.
Equivalently, one may use a form akin to that developed by Fano \cite{Fan61} 
and Fano and Cooper \cite{Fan65} 
to describe the Beutler-Fano profiles observed in autoionization or
electron-scattering cross-sections, 
\begin{equation}  \label{Fanoprofile}
\sigma(\omega) = \sigma_b(\omega) + \sigma_a \frac{ (x + q)^2} {x^2 + 1}
\end{equation}
where $q$ is the dimensionless Fano $q$ parameter. The terms $C(\omega)$ or $%
\sigma_b(\omega)$ are slowly varying (or constant) contributions to the
cross section. Appendix A describes simple estimators of the parameters when
fitting experimental data to this functional form.

We found it useful to use a variant of these parameterizations, 
\begin{equation}  \label{ourprofile}
\sigma(D) = \sigma_a \mbox{Re} \left[ 1 - \frac{ (1 - iq)^2\Gamma_d}{%
\Gamma_d + \Gamma_0 - i2(D - D_0)} \right]
\end{equation}
where $D$ is the detuning of the probe laser, and $\Gamma_d$ and $\Gamma_0$
are parameters discussed in section \ref{secFit}.

Although we have presented profile formulas appropriate to photoionization
cross-sections, what is important for the present paper is the functional
dependence of an experimentally observed quantity, here $\sigma(\omega)$,
upon a dimensionless detuning variable $x$. It often proves convenient to
fit dimensionless observables with these parameterizations, in which case
the parameters $A$, $B$ and $C$, or $\sigma_a$ and $\sigma_b$, become
dimensionless.

\subsection{Present work}

In an earlier paper \cite{Hal98} 
we presented experimental evidence for substantial suppression of
photoionization by means of LICS. As we stressed, only with the use of very
coherent laser pulses, together with an appropriate choice of atomic states,
is it possible to observe pronounced suppression. Here we present further
experimental results showing dependence of the LICS profile on various pulse
properties.

Our experimental results are for LICS induced in metastable helium, by means
of a dressing laser (1064 nm) that couples the {4s ${}^1$S$_0$} state into a
continuum. The structure is observed by scanning the frequency of a probe
laser which, at a wavelength of 294 nm, couples the {2s ${}^1$S$_0$} state
into this same continuum. {Figure \ref{fig. 3}} shows the relevant energy
levels and transitions.

We discuss numerical simulation of the experimental results, and show that
with calculations of atomic parameters the observations are in excellent
agreement with computations. We present an analytic description of pulsed
LICS, emphasizing the difference between pulsed and steady excitation. These
results reduce to expressions (\ref{BWprofile}) and (\ref{Fanoprofile})
under appropriate conditions. Our calculations of needed atomic parameters
(polarizabilities, Stark shifts, etc.) are done using a simple
effective-potential model of the active electron (see Appendix \ref{model}).
We describe this method, and we show that this model provides quantitative
agreement with experiments.

\newpage

\section{Experimental}

Our observations of LICS in the photoionization of helium are carried out
using an atomic beam of metastable helium atoms, which cross spatially
overlapping beams of pulsed laser radiation. We control the duration of the
pulses and the overlap in space and time. {Figure \ref{figlayout}} shows a
schematic diagram of the experimental apparatus.

\subsection{The apparatus}

\paragraph{The atomic beam.}

A pulsed beam of metastable helium atoms is formed as helium, at a
stagnation pressure of 1200 mbar, expands through a nozzle (General Valve,
opening diameter 0.8 mm) and enters a region of pulsed discharge located 4
mm from the nozzle. A skimmer (diameter 0.8 mm) placed 40 mm from the nozzle
collimates the atomic beam and separates the source chamber from the region
of interaction and detection.

\paragraph{The metastable atoms.}

The metastable states of helium used for our photoionization experiments are
populated in a gas discharge, where the atoms are excited from the ground
state {1s ${}^1$S$_0$} by pulsed electron impact. To enhance efficiency and
stability of the discharge, a filament, placed 10 mm downstream, provides
the discharge with a continuous electron current of several mA. The peak
current, produced by ions and electrons in the pulsed discharge, reaches 20
mA. The electron source enhances the efficiency of the metastable production
by at least two orders of magnitude and the breakdown voltage is reduced to
1/50 of the value needed for an unseeded discharge.

\paragraph{Diagnostics of metastable population}

We deduce the population of metastable atoms by using a two-photon version
of REMPI (resonant-enhanced multiphoton ionization) to produce photoions,
with pulsed radiation from a broadband dye laser (LPD 3000, Lambda Physik).
The ions are observed as an electric current. {Figure \ref{fig. 2}} shows
the REMPI transitions used to detect the {2s ${}^1$S$_0$} and {2s ${}^3$S$_1 
$} metastable atoms.

{Figure \ref{fig. 1}} shows the ionization signal of the metastable singlet
and triplet states as a function of probe-laser wavelength. The REMPI
transitions are strongly broadened by saturation. We also confirmed that the
ionization is saturated. Populations are inferred from the heights of the
ionization signal peaks. Typically 90\% of the metastable population is
found in the triplet state {2s ${}^3$S$_1$} , and 10\% is found in the
singlet state {2s ${}^1$S$_0$} .

\paragraph{The lasers.}

The pulsed atomic beam crosses the coinciding axes of two pulsed laser
beams: a UV probe pulse and an IR dressing pulse. 
The probe laser field, coupling the metastable state {2s ${}^1$S$_0$} to the
ionization continuum (see {Fig. \ref{fig. 3}}), is derived from a single
mode cw dye laser system operating at 587 nm. This radiation is amplified in
a pulsed dye amplifier, pumped by the second harmonic of an injection-seeded
Nd:YAG laser. The wavelength of the cw radiation is measured to an accuracy
of $\Delta\lambda / \lambda = 2 \times 10^{-6}$ in a Michelson-type
wavemeter, using a He-Ne laser, stabilized on an iodine line, as a
reference. The pulsed radiation is frequency-doubled in a BBO crystal,
thereby providing (probe) laser pulses with a pulse width $\Delta \tau = 2.3$
ns (half-width at $1/e$ of $I(t)$) at $\lambda = $ 294 nm.

The pulse of the dressing laser couples the state $4s^1S_0$ to the
ionization continuum and induces the continuum structure. This pulse,
obtained from the fundamental frequency of the Nd:YAG laser ($\lambda =$
1064 nm), has a pulse width $\Delta \tau = 5.1$ ns.

A folded optical delay line is used to adjust the time delay between the
dressing and probe pulses. After passing through the optical setup, pulse
energies of up to 2 mJ for the probe laser and up to 250 mJ for the dressing
laser are delivered to the atomic beam. The laser diameters at the atomic
beam position are 0.5 mm for the probe and 3.5 mm for the dressing laser,
yielding peak intensities of up to $I_p^{(0)} =$ 100 MW/cm$^{2}$ and $%
I_d^{(0)} =$ 300 MW/cm$^{2}$, respectively. (we use a superscript $I^{(0)}$
to denote the peak value of a time-varying pulse $I(t)$.) Because of the
large diameter of the dressing-laser beam, the variation of its intensity
across the probe laser profile is small.

\paragraph{Ionization detection.}

Ionized helium atoms are mass-selectively detected with a double-thickness
micro-sphere plate (El Mul Technologies) after passing through a 30 cm
time-of-flight segment. The output current of the micro-sphere plate is
amplified with fast broadband amplifiers and integrated in a boxcar gated
integrator (EG \& G 4121 B).

\subsection{Results}


The continuum structure induced by the dressing laser is revealed in the
ionization of the metastable state {2s ${}^1$S$_0$} produced by the probe
laser. When the probe laser is tuned across the two-photon resonance between
the states {2s ${}^1$S$_0$} and \mbox{4s ${}^1$S$_0$}, we observe a strong
and spectrally narrow feature in the ionization cross-section (see the
following subsections). The structure is due to the laser-induced mixing of
the bound state {4s ${}^1$S$_0$} with the ionization continuum.

\subsubsection{Background signal}

At high laser intensities, two-photon ionization of the metastable triplet
state (see {Fig. \ref{fig. 3}}) forms a background, which has to be
subtracted from the total ion signal. The background ionization was
determined by tuning the UV laser away from the region where LICS is
observed and measuring the ionization probability with and without the IR
laser present. The difference of these two currents (the difference signal)
is attributed to the two-photon (IR + UV) ionization of the triplet level,
the population of which exceeds that of the singlet level by about an order
of magnitude.

{Figure \ref{fig. 4}} shows the measured difference signal, relative to the
total ion signal, as a function of probe intensity. The probe laser
frequency is here tuned also far away from the two photon resonance between
states {2s ${}^1$S$_0$} and \mbox{4s ${}^1$S$_0$}. The plotted signal is
proportional to the ionization probability of the {2s ${}^3$S$_1$} state.
For comparison we show the calculated ionization probability for {2s ${}^3$S$%
_1$} state caused by one photon at 294 nm and one photon at 1064 nm, using
the theoretical ionization probability. Using wavefunctions described in
Appendix \ref{model} we obtained the result $\Gamma_{triplet} = 1.6 \times
10^{-9} I_p I_d$. Here and elsewhere laser intensities are expressed in W/cm$%
^2$ and the unit for the rate is s$^{-1}$. The calculated relative
transition probability is in good agreement with the experimental data.

From {Fig. \ref{fig. 4}} we see that the background signal contributes only
a few percent to the total ion signal for $I^{(0)}_p < 10$ MW/cm$^2$, while
it adds more than 30\% for the highest probe laser intensity used here. This
background signal has been subtracted for all the data presented below.

\subsubsection{LICS for weak probe laser}

If the probe laser is weak, the shape of the continuum structure is
dominated by the strong dressing laser, and the probe does not affect the
internal structure of the atom. It is therefore possible to distinguish
between the influence of the dressing and probing lasers.

{Figure \ref{fig. 5}} shows, for a weak probe laser, the observed structure
in the ionization continuum of helium for various dressing-laser intensities
(in each column) and for two pulse timings (in each row). The left-hand
column of frames show the LICS produced by coincident laser pulses; the
right-hand frames show LICS produced when the dressing pulse precedes the
probe pulse (negative delay) by 5 ns. 
The dotted lines are results from numerical simulations, discussed in
section \ref{simulation}.

The continuum resonances exhibit typical asymmetric window-resonance
structure as a function of frequency, such as are observed in
autoionization. The LICS exhibit an enhancement of photoionization, taking
largest value $P_{max}$ at frequency $\nu_{max}$, and a suppression of
photoionization, a dip to $P_{min}$ centered at frequency $\nu_{min}$.

As can be seen, it is possible to control the photoionization probability by
appropriately tuning the wavelength of the probe laser, thereby selecting
the region of enhancement or suppression of ionization. Photoionization
suppression indicates population trapping. We observe photoionization
suppression as strong as a drop to 30\% of the value far from resonance \cite
{Hal98}.

The primary effects of an increase in dressing-laser power are a broadening
and a shift of the LICS profile, both for coincident and delayed laser
pulses. This is expected: the dressing laser intensity is responsible for
the width-parameter $\Gamma$ and the Stark shifts.

For delayed laser pulses the LICS is not as strong as, and is spectrally
narrower than, for coincident pulses. This is because when pulses are not
coincident, the dressing laser is intensity is lower when the probe pulse
samples the continuum structure: the LICS width is smaller and less
pronounced when the dressing is weaker.

\subsubsection{LICS for strong probe laser}

For a strong probe laser, it is no longer possible to distinguish between
the effect of the dressing and probing laser pulses. Both pulses contribute
to the continuum structure. Whether the influence of the probe pulse is
negligible or not can be estimated from the ionization probability of state %
\mbox{2s ${}^1$S$_0$}. {Figure \ref{fig. 6}} shows the measured ionization
probability of state {2s ${}^1$S$_0$} as a function of the probe laser
intensity (in the absence of the dressing laser). The observed dependance is
in good agreement with the theoretical prediction ({Fig. \ref{fig. 6}},
dashed curve), based on the calculated ionization rate $\Gamma_{2s,p}$ (see
Table \ref{ratetable}).

For probe intensities of several tens of MW/cm$^2$ the ionization is
strongly saturated. In this regime the interaction induced by the probe
laser cannot be considered as a small perturbation of the atomic structure.
The observed profile will consequently deviate from the profiles of eqns. (%
\ref{BWprofile}) and (\ref{Fanoprofile}).

{Figure \ref{fig. 7}} compares the observed LICS for weak and strong probe
pulses. The peak values of probe laser intensities are $I_p^{(0)} = 0.5 $
MW/cm$^2$ ({Fig. \ref{fig. 7}}a) and $I_p^{(0)} = 40 $ MW/cm$^2$ ({Fig. \ref
{fig. 7}}b). The peak intensity of the dressing laser was $I_d^{(0)} = $ 44
MW/cm$^2$ and 50 MW/cm$^2$. An increase of the probe laser intensity reduces
the peak value of the ionization, relative to the value far from resonance,
from 140\% to 110\%. This reduction of the photoionization enhancement is
the result of saturation: the atoms are rapidly photoionized for all
frequencies except those where photoionization is suppressed. Because the
strong probe pulse ionizes about 90\% of the population of the metastable
singlet state (see {Fig. \ref{fig. 6}}), the maximum enhancement of the
ionization probability can be no more than 10\%.

\subsection{Fitting Experimental Data}

\label{secFit}

To parameterize our experimental results we first subtracted a linearly
varying background fitted to the far-from-resonance values, and then
performed a least-squares fit of the resulting data to a version of the
Breit-Wigner or Fano-Cooper formula, Eqn. (\ref{ourprofile}). This
parameterization introduces two contributions to the overall width parameter 
$\Gamma = \Gamma_d + \Gamma_0$. The ionization rate of the dressing laser, $%
\Gamma_d$, can, in principle, be computed using the traditional Fermi Golden
Rule. The remaining term, $\Gamma_0$, represents an additional empirical
rate; it expresses the effect of spontaneous emission, of fluctuations of
intensity or phase, and other stochastic processes. {Figure \ref{fittingfig}}
shows an example of the experimental data (small squares) and the best
analytic fit to this curve (dashed line). As can be seen, the data is very
well described by the analytic expression. Table \ref{table1} presents
values of the least squares parameters for undelayed pulses.

\begin{minipage}{4in}
\begin{table}[h]
\caption{Fits of the experiment, $\Delta t = 0$}
\begin{tabular}{ccccccccc}
$I_p^{(0)}$  & $I_d^{(0)}$  & $q $  & $\Gamma$ & $\Gamma_0$ & $\Gamma_d$
& $\Gamma_{theory}$
\\
\hline
\\
2.5 & 38 & 0.66   & 0.83     & 0.29 & 0.54 & 0.45
\\
4 & 75 & 0.71   & 1.36     & 0.42 & 0.94  & 0.88
\\
4 & 140 & 0.67    & 2.21     & 0.61 & 1.61 & 1.64
\end{tabular}
\label{table1}
\end{table}
Here $I^{(0)}_p$ and $I^{(0)}_d$ are the peak intensities of the probe and
dressing lasers
in MW/cm$^2$, $q$ is the Fano $q$ parameter, $\Gamma_d$ is the ionization rate
from the dressing laser, $\Gamma$ is the overall width parameter,
$\Gamma_0 = \Gamma - \Gamma_d$, and $\Gamma_{theory}$ is the theoretical
value for $\Gamma_d$. All rates are in units of GHz. \\
\end{minipage}

From several such fits we found the trend of the width parameters as a
function of dressing intensity. {Figure \ref{widthvsI}} plots values of the
parameter $\Gamma_0$ and of the empirically determined $\Gamma_d$. We show
also the theoretical value $\Gamma_{theory}$ of the photoionization rate $%
\Gamma_d$, calculated using the method described in section \ref{model}.
Theory and experiment are in very good agreement.

Laser intensity fluctuations lead to fluctuations in the laser-induced Stark
shifts. These fluctuations give an additional contribution to the LICS
width, 
\begin{equation}
\Gamma_0^{Stark} = \langle S_{4s,d}^0 - S_{2s,d}^0 \rangle_{Fluc}
\end{equation}
Here $S_{i,j}^0$ is the peak Stark shifts of state $i$ induced by laser
field $j$ (see Section \ref{simulation}). With $I_d^{(0)} = 75$ MW/cm$^2$
and typical laser intensity fluctuations of $\Delta I_d^{(0)} = \pm$ 10 \% $%
I_d$ the additional broadening is $\Gamma_0^{Stark}/2\pi = 0.2 $ GHz.

\newpage

\section{Numerical simulation}

\label{simulation}

Usually laser induced continuum structures are deduced from the equations
for the density matrix of an atom. This approach allows one to take into
account coherence dephasing processes such as laser frequency fluctuations,
collisions etc. But in the case of coherent laser pulses, with
transform-limited bandwidths, one can significantly simplify the description
of the process by using the time-dependent Schr\"odinger equation for the
amplitudes of atomic states involved in the process.

\subsection{The main equations}

We consider an atom interacting with two laser fields: a probe laser with
amplitude ${\cal E}_p(t)$, intensity $I_p(t)$ and carrier frequency $\omega
_p$, and a dressing laser with amplitude ${\cal E}_d(t)$, intensity $I_d(t)$
and carrier frequency $\omega _d$. The total electric field is 
\begin{equation}  \label{eq2}
{\bf E}(t) = \mbox{Re}\sum_{j=p,d}{\bf e}_j{\cal E}_j(t) \exp \left(
-i\omega_jt\right),
\end{equation}
where ${\bf e}_j$ is a unit vector. The probe and dressing lasers are tuned
close to the two-photon Raman resonance between states 1 and 2, and their
carrier frequencies differ from exact resonance by the two-photon detuning
D, given by 
\begin{equation}  \label{eq3}
\hbar D \equiv \hbar (\omega _p-\omega _d) -(E_2-E_1).
\end{equation}

We express the state vector $\left| \Psi \left( t\right) \right\rangle \ $
of the atom as a combination of two particular discrete states 1 and 2, with
energies $E_1$ and $E_2$, together with a sum over all other bound states
and an integral over continuum states: 
\begin{eqnarray}  \label{eq1}
|\Psi(t)\rangle & = & \sum\limits_{n=1,2} C_n(t) \exp \left( -i
E_nt/\hbar\right) |n\rangle \\
& + & \sum\limits_{n\neq 1,2} A_n(t) \exp \left( -i E_nt/\hbar \right)
|n\rangle \\
& + & \sum\limits_j\int dE\, A_{E,j}(t) \exp \left( -i Et/\hbar \right)
|E,j\rangle,
\end{eqnarray}
where $|E,j\rangle$ is the continuum state with energy $E$ (quantum number $%
j $ identifies different possible continua).

As is customary \cite{Den85a,Yat97,Sho90,Kni90,Cra80,Dai87,Nak94} 
we adiabatically eliminate all amplitudes except those of states 1 and 2. In
so doing, we obtain effective interactions, and dynamic energy shifts,
expressible in terms of a polarizability tensor \cite{Sho90}, 
\begin{equation}
{\sf Q}_{mn}(\omega )={\sum_{k}}\frac{\langle m|{\bf d}|k\rangle \langle k|%
{\bf d}|n\rangle }{E_{k}-E_{m}-\hbar \omega }+\int dE{\sum_{j}}\frac{\langle
m|{\bf d}|E,j\rangle \langle E,j|{\bf d}|n\rangle }{E-E_{m}-\hbar \omega },
\label{polar}
\end{equation}
where ${\bf d}$ is the dipole-moment operator. The continuum integral
includes the resonance energy $E=E_{m}+\hbar \omega $, which is dealt with
by breaking the integration into a nonsingular principal value part (denoted
by ${\cal P}$) and a resonant part (yielding the transition rate of a Fermi
Golden Rule), 
\begin{eqnarray}
\mbox{Re}{\sf Q}_{mn}(\omega ) &=&{\cal P}{\sum_{k}}\frac{\langle m|{\bf d}%
|k\rangle \langle k|{\bf d}|n\rangle }{E_{k}-E_{m}-\hbar \omega }+\int dE{%
\sum_{j}}\frac{\langle m|{\bf d}|E,j\rangle \langle E,j|{\bf d}|n\rangle }{%
E-E_{m}-\hbar \omega },  \nonumber \\
\mbox{Im}{\sf Q}_{mn}(\omega ) &=&\sum_{j}\pi \langle m|{\bf d}%
|E=E_{m}+\hbar \omega ,j\rangle \langle E=E_{m}+\hbar \omega ,j|{\bf d}%
|n\rangle .
\end{eqnarray}
The adiabatic elimination yields a two-state time-evolution equation, which
may be written \cite{Kni90} 
\begin{equation}  \label{eq4}
i\frac{d}{dt}\left[ 
\begin{array}{c}
C_{1}(t) \\ 
C_{2}(t)
\end{array}
\right] =\frac{1}{2}\left[ 
\begin{array}{rl}
2S_{1}(t)-i\Gamma _{1}(t)-i\gamma _{1}\quad & -(i+q)\Omega (t) \\ 
-(i+q)\Omega (t){\quad } & 2S_{2}(t)-i\Gamma _{2}(t)-i\gamma _{2}-2D
\end{array}
\right] \left[ 
\begin{array}{c}
C_{1}(t) \\ 
C_{2}(t)
\end{array}
\right] .  \label{eqn12}
\end{equation}
This equation, the basis for our simulation, requires dynamic Stark shifts $%
S_{i}(t)$, detuning $D$, photoionization rates $\Gamma _{i}(t)$, spontaneous
emission rates $\gamma _{i}$, an effective Rabi frequency $\Omega (t)$,
defined below in Eqn. (\ref{effectiverabi}), and the Fano parameter $q$.
These various parameters can be obtained from atomic wavefunctions, as the
following paragraphs show.

We are interested in the total ionization probability after pulses cease, $P$%
. This can be determined if we know the populations of states {1} and {2}
after the action of the laser pulses: 
\begin{equation}  \label{eq8}
P=1-\left| C_1\left( \infty \right) \right| ^2 -\left| C_2\left(\infty
\right) \right| ^2.
\end{equation}
The LICS profile is taken as the variation of $P$ with two-photon detuning $%
D $.

\subsection{Formulas for Atomic Parameters}

The theoretical ionization rates and Stark shifts are obtainable by summing
partial rates: 
\begin{equation}
\Gamma _{i}(t)=\sum\limits_{j,\alpha }\Gamma _{i\alpha }^{(j)}(t),\qquad
S_{i}(t)=\sum\limits_{\alpha }S_{i\alpha }(t),  \label{sumrates}
\end{equation}
where $\Gamma _{i\alpha }^{(j)}(t)$ is the ionization rate from state $i$ to
continuum $j$ caused by probe ($\alpha =p$) or dressing ($\alpha =d$) lasers 
\begin{equation}
\Gamma _{i\alpha }^{(j)}(t)=\frac{\pi }{2\hbar }\left| {\cal E}_{\alpha
}(t)\right| ^{2}\left| \langle i|{\bf e}_{\alpha }{\cdot {\bf d}}|E=\hbar
\omega _{\alpha }-E_{i},j\rangle \right| ^{2},  \label{ionrate}
\end{equation}
and $\hbar S_{i\alpha }(t)$ is the dynamic Stark shift of the energy of
state $i$ produced by laser $\alpha $: 
\begin{equation}
S_{i\alpha }(t)=-\frac{1}{4\hbar }\left| {\cal E}_{\alpha }(t)\right| ^{2}%
{\bf e}_{\alpha }{\bf \cdot }\left[ {\sf Q}_{ii}(\omega _{\alpha })+{\sf Q}%
_{ii}(-\omega _{\alpha })\right] {\bf \cdot e}_{\alpha }^{*}.
\label{starkshift}
\end{equation}
The quantity 
\begin{equation}  \label{rabifreq}
\Omega (t)=\sum\limits_{j}\sqrt{\Gamma _{1p}^{(j)}(t)\Gamma _{2d}^{(j)}(t)}
\label{effectiverabi}
\end{equation}
is an effective Rabi frequency for the two-step transition from the state 1
to the state 2 with intermediate population of the continuum states $j$ with
energy $E\simeq \hbar \omega _{p}-E_{2}+E_{1}\simeq \hbar \omega
_{d}-E_{1}+E_{2}$. The Fano $q$ parameter can be evaluated from the
expression \cite{Yat97,Kni90,Dai87,Nak94} 
\begin{equation}
q\Omega (t)=\frac{1}{2\hbar }\left| {\cal E}_{p}(t){\cal E}_{d}(t)\right| 
{\bf e}_{p}{\bf \cdot }\left[ {\sf Q}_{12}(\omega _{p})+{\sf Q}_{12}(-\omega
_{d})\right] {\bf \cdot e}_{d}^{*}.  \label{eq7}
\end{equation}
From Eqns. (\ref{effectiverabi}) and (\ref{eq7}) it is apparent that $q$
depends only on atomic properties.

In principle, all of these quantities can be calculated if the atomic
wavefunctions are known.

We note that the Fano parameter $q$ is basically the ratio of polarizability
(such as occurs in a Raman transition) to the product of two dipole
transition moments into the continuum. In systems with $|q| \gg 1$ the
Raman-type transitions dominate. The ionization probability is enhanced,
when tuning the probe laser across the two-photon resonance, because
Raman-type transitions open additional (multiphoton) ionization channels
from the initial state to the ionization continuum. If $q = 0$ no Raman-type
transitions are present and no enhancement of ionization can be observed.

\subsection{Values for Atomic Parameters}

For the calculations we need the values of many transition matrix elements,
both bound-bound and bound-continuum. Due to the relative simplicity of the
two-electron helium atom there exist several methods, of varying complexity
and accuracy, which can provide the needed wavefunctions \cite
{Bur65,Sch71,Lan77,Kon84,Che94}. 
However, these methods are usually optimized for the calculation of
particular parameters (such as the ionization energy).

Therefore, for our purposes it is desirable to use some simpler method. The
simplest of these are based on the assumption that each electron moves in an
effective central potential, and that therefore the atomic wavefunction is
separable into a product (antisymmetrized) of individual electron orbitals,
each of which has a spherical harmonic for angular dependence and a radial
dependence given by the solution to an ordinary differential equation (see
Appendix \ref{model}).

We have found that a simple single-electron model potential satisfies
requirements for speed and accuracy. Appendix \ref{model} describes the
mathematics of this method for computing matrix elements. As noted there, a
single empirically determined parameter $\lambda_{\ell}$ fixes the
potential, and hence all the wavefunctions, for each orbital angular
momentum $\ell$.

Table \ref{ratetable} presents the theoretical values we used in the
simulation.

\begin{minipage}{4in}  

\begin{table}[h]
\caption{Theoretical values for simulation}
\label{ratetable}
\begin{tabular}{cccc}
$q$ & 0.73
\\
$\Gamma _{1p}$ & $13.4\times I_p$
&
 $S_{1p}$ & $9.0\times I_p$
\\
$\Gamma _{2p}$ & $1.9\times I_p$
&
$S_{2p}$ & $13.0\times I_p$
\\
 $\Gamma _{1d}$ & $0$
&
 $S_{1d}$ & $70.0\times I_d$
\\
$\Gamma _{2d}$ & $73.7\times I_d$
&
$S_{2d}$ & $142.0\times I_d$
\end{tabular}
\end{table}
\end{minipage}

Here $I_p$ and $I_d$ are the intensity of probe and dressing lasers
expressed in W/cm$^2$. The ionization rates and Stark shifts are expressed
in s$^{-1}$. These results are in good agreement with more elaborate
computations carried out by Klystra, Paspalakis and Knight \cite{Kly98}. 
The agreement with these more accurate results, and the successful fitting
of our experimental data, support the validity of our atomic wavefunctions.

The numerical simulation of the observed LICS-profile includes
experimentally determined intensity fluctuations and integration over the
spatial probe laser intensity profile. These effects have to be taken into
account, because position, bandwidth and shape of the LICS are strongly
dependent on the laser power.

To account for pulse-to-pulse variations of the intensity we average our
computed ionization probabilities over a distribution of peak pulse
intensities. We used a Monte-Carlo method to estimate the relevant
integrals, assuming intensities evenly distributed between 0.9 $I_{p}$ and
1.1 $I_{p}$. The results of the simulations are in excellent agreement with
the experimental results, as are seen in {Fig. \ref{fig. 5}}.

\subsection{Deviations from Steady State}

\label{sec_deviations}

When the probe pulse is weak, and much shorter than the strong dressing
pulse, the excitation can be regarded, to a first approximation, as an
example of steady-state LICS 
(see sections \ref{sec_quasistationary} - \ref{sec_quasistationary2}). One
expects, under these circumstances, that an observed LICS profile will
follow the form of Eqns. (\ref{BWprofile}) or (\ref{Fanoprofile}) \cite
{Kni90} or, equivalently (\ref{ourprofile}), and that the parameters $%
q_{fit} $ and $\Gamma_{fit}$ obtained from fitting such an expression will
be equal to the values $q_{theo} $ and $\Gamma_{theo}$ obtained by
evaluating matrix elements. 

To show the effects of pulse shapes we display, in {Fig. \ref{twoapprox}}
the profiles predicted in two approximations:

\begin{itemize}
\item  Steady-state profile, using the theoretical $q$ value.

\item  Simulation for a single pair of pulses (i.e. transient theory), again
using the theoretical $q$ value.
\end{itemize}

The difference between these curves is small but it shows the importance of
transient effects and the limitation of the steady-state approximation of
sections \ref{sec_quasistationary} - \ref{sec_quasistationary2}. 
Here ``transient'' means that we determine the ionization probability from
the solution of the differential equations (\ref{eq4}) thereby taking into
account the pulsed variability of the probe interaction. The spectral width
of this pulse, approximately $1/\tau_p$, effectively contributes additional
broadening, $\Gamma_0$, to the LICS profile. 
Only with the steady-state approximation ($\tau_p \rightarrow \infty$ and $%
\Gamma_0=0$) is there complete suppression of photoionization.

\subsection{Delayed Pulses}

In addition to the results just described, for which the probe pulse was
timed to coincide with the peak value of the dressing pulse, we have also
obtained results with probe pulses delayed with respect to the peak of the
dressing pulse. As with the previously discussed results, it is possible to
fit these LICS profiles with the function shown in Eqn. (\ref{ourprofile}).
Table \ref{table2} gives the values of parameters from these fits.

\begin{minipage}{4in}
\begin{table}[h]
\caption{Fits of experiment, $\Delta t = -5$ ns}
\begin{tabular}{ccccccccc}
$I_p^{(0)}$  & $I_d^{(0)}$  & $q $  & $\Gamma$ & $\Gamma_0$ & $\Gamma_d$
& $I_d^{(eff.)}$
\\
\hline
\\
2.5 & 38 & 0.63    & 0.66   & 0.32 & 0.34 & 29
\\
4 & 75 & 0.67    & 0.97     & 0.42 & 0.55 & 47
\\
4 & 140 & 0.67    & 1.72    & 0.57 & 1.15 & 98
\end{tabular}
\end{table}
\label{table2}
\end{minipage}

The width parameters obtained from fits for delayed pulses are significantly
larger than what one would expect from the value of the dressing intensity
at the peak of the probe pulse. To emphasize this point, {Fig. \ref
{delayedprobe}} shows the dressing and probe envelopes. The point marked B
indicates the dressing intensity required to produce the observed width. The
point marked A shows the (lower) intensity occurring at the peak of the
probe pulse. 
As these points show, it is not possible to evaluate the LICS profile by
considering only the moment of peak probe intensity. The observed profile
averages in time the Stark shifts and the ionization rates.

\subsection{Summary}

In the simulations of the experimental data presented above, intensity
fluctuations of both lasers as well as an integration over the spatial probe
laser profile are taken into account to describe the observed LICS. In the
present experiments the diameter of the dressing laser was much larger than
that of the probe laser, and so the spatial variation of the dressing laser
intensity was neglected in the simulation.

\paragraph{Weak probe}

When the probe laser is weak, the continuum resonance is mainly determined
by properties of the dressing laser. Fluctuations in the dressing laser
intensity can disturb the observation of the continuum resonance. A change
in the dressing laser intensity changes the Stark shifts as well as the peak
value and central wavelength of the LICS. Because the Stark shifts follow
the fluctuations of the dressing laser intensity, the position of the
structure varies with every laser pulse. The observed LICS is therefore
broadened by the fluctuations. In the most extreme cases the detected ion
signal, for a fixed detuning, fluctuates between the maximum ionization
enhancement and the ionization suppression. The averaging produced by the
fluctuation reduces the LICS contrast between maximum enhancement and
minimum suppression.

A weak probe laser does not affect the LICS, and so the probe laser
intensity fluctuations have no influence on the continuum resonance if there
are no dramatic variations in the laser intensity for single pulses. It is
not necessary, in simulations, to 
average over intensity fluctuations of a weak probe laser.

\paragraph{Strong probe}

When the probe laser is strong the continuum is modified by the dressing
laser {\it and} by the probe laser. The influence of the strong probe laser
can be observed in the saturation of ionization, i.e. a reduced enhancement
in the continuum resonance. Although saturation might be reached in the
center of the spatial probe laser profile, the intensity could still be
below saturation in the wings. The maximum enhancement is stronger in the
wings than in the center of the profile. An integration over the spatial
profile of a strong probe laser therefore leads to less saturation than
expected for constant strong laser intensity.

Fluctuations in the probe laser intensity determine, as do fluctuations in
the dressing laser intensity, additional Stark shifts, i.e. variations in
linewidth and position of the continuum resonance, as discussed above.



\section{Analytical expressions for the LICS profile}

It is always instructive to augment numerical simulation with analyses that
can provide explicit formulas for various observables. We present in the
following sections some examples of such formulas. We consider pulsed
excitation, so that one cannot always rely on the stationary solutions that
are known to give the formulas Eqn. (\ref{BWprofile}) and Eqn. (\ref
{Fanoprofile}) for the LICS shape.

\subsection{General theory}

If we neglect the coupling $\Omega(t)$ of Eqn. (\ref{eq4}), then the
amplitude of $C_1(t)$ decays exponentially with time, while its phase
depends on the Stark shift $S_1(t)$. With this in mind we introduce the new
variables 
\begin{eqnarray}  \label{eq8a}
h_1(t) &= & \ln C_1(t) -i\int_{-\infty }^tdt^{\prime }\,S_1(t^{\prime }), \\
h_2(t) &=&C_2(t)/C_1(t)
\end{eqnarray}
In terms of the new variables the populations of states {1} and {2} after
the interaction with the laser pulses are 
\begin{equation}
| C_1( t )| ^2 =\exp[2 {\rm Re} h_1(t)] \quad \mbox{ and } \quad | C_2( t )|
^2 =\exp[2 {\rm Re}h_1(t)] \, |h_2(t)|^2.
\end{equation}
In the following we assume the probe pulse ${\cal E}_p(t)$ to be weak,
meaning that for the pulse duration $\tau _p$ the direct ionization
probability 
\begin{equation}
\varepsilon ^2 \equiv \int_{\infty}^{\infty} dt
\sum\limits_j\Gamma_{1p}^{(j)}(t)
\end{equation}
remains small: $\varepsilon ^2 \ll 1$. We therefore seek solutions in the
form of a power series in the parameter $\varepsilon $. Because for $%
\varepsilon =0$ the populations are $| C_1( t )| ^2=1$ and $| C_2( t )|^2=0$%
, the variables $h_i(t)$ vanish for $\varepsilon =0$ and the series
expansion reads 
\[
h_i(t) = \sum_k \varepsilon^k h_i^{(k)}(t). 
\]

We assume that the state {1} is long lived so that one can neglect the
spontaneous decay during the interaction with the laser pulses ($\gamma _1=0$%
). We assume also that the frequency $\omega _2$ is small enough that the
dressing field does not ionize the atom in state {1}, and hence $\Gamma
_1(t)=\sum\limits_j\Gamma _{1p}^{(j)}(t)$. The equations for $h_1(t)$ and $%
h_2(t)$ derived from Eqn. (\ref{eq4}) then read 
\begin{equation}  \label{eq9}
\frac{ d}{dt}h_1(t)=-\frac{\Gamma _1(t)}2+\frac{(iq-1)}2\Omega(t) h_2(t),
\end{equation}
\begin{equation}  \label{eq10}
\frac{ d}{dt}h_2(t)=-\left( \frac{\Gamma _2(t)}{2} +\frac{\gamma _2}{2}-%
\frac{\Gamma _{1p}(t)}{2} -iD_{eff}(t)\right) h_2 (t)+\frac{(iq-1)}{2}%
\Omega(t) (1-h_2(t)^2),
\end{equation}
where $D_{eff}(t)=D-S_2(t)+S_1(t)$ is the two-photon detuning incorporating
the Stark-shifted resonance. We require solutions that vanish initially: $%
h_1(-\infty )=h_2(-\infty )=0$.


The value of $h_{1}(t)$ obtained from Eqn. (\ref{eq9}), when integrated over
the pulse duration $\tau _{p}$, is proportional to $\varepsilon ^{2}$. To
first order in $\varepsilon $ Eqn. (\ref{eq10}) reads 
\begin{equation}
\frac{d}{dt}h_{2}^{(1)}(t)=-\left( \frac{\Gamma _{2}(t)}{2}+\frac{\gamma _{2}%
}{2}-iD_{eff}(t)\right) h_{2}^{(1)}(t)+\frac{(iq-1)}{2}\Omega (t).
\label{eq11}
\end{equation}
The solution of this equation with the initial condition $h_{2}(-\infty )=0$
is 
\begin{equation}
h_{2}^{(1)}(t)=\frac{(iq-1)}{2}\int_{0}^{\infty }d\tau \,\Omega (t-\tau
)\exp \left[ -\int_{t-\tau }^{t}dt^{\prime }\,\left( \frac{\Gamma
_{2}(t^{\prime })}{2}+\frac{\gamma _{2}}{2}-iD_{eff}(t^{\prime })\right)
\right] .  \label{eq12}
\end{equation}
Using Eqns. (\ref{eq12}) and (\ref{eq9}) we find $h_{1}^{(2)}(t)$ in second
order of perturbation theory to be: 
\begin{equation}
h_{1}^{(2)}(t)=-\int\limits_{-\infty }^{t}\frac{\Gamma _{1}(t^{\prime })}{2}%
dt^{\prime }+\frac{(iq-1)}{4}\int\limits_{-\infty }^{t}\Omega (t^{\prime
})h_{2}^{(1)}(t^{\prime })dt^{\prime }.  \label{eq13}
\end{equation}
From (\ref{eq12}) one can see that $h_{2}^{(1)}(+\infty )=0$ and hence, to
second order in the small parameter $\varepsilon $, the ionization
probability $P$ is 
\begin{equation}
P=1-\exp (-\kappa ),  \label{eq14}
\end{equation}
where the exponent $\kappa $ is the value of the variable $h_{1}(t)$ in the
limit of long times, 
\[
\kappa =-2\mbox{Re}\left[ h_{1}^{(2)}(+\infty )\right] . 
\]

Using Eqn. (\ref{eq13}) for $h_1^{(2)}(t)$ and Eqn. (\ref{eq12}) for $%
h_2^{(1)}(t) $ we can write the parameter $\kappa $ as 
\begin{eqnarray}  \label{eq15}
\kappa &= & \kappa_0 - \mbox{Re}\left\{ \frac{(iq-1)^2}{2}
\int\limits_{-\infty }^{+\infty} dt \int\limits_0^\infty d\tau \, \Omega
(t)\,\Omega (t-\tau ) \exp \left[-\varphi(t,\tau) \right] \right\},
\end{eqnarray}
where $\kappa_0$ is the value of the parameter $\kappa$ in the absence of
the dressing laser, 
\begin{equation}
\kappa_0 \equiv \int\limits_{-\infty }^{+\infty }\Gamma _1(t)\,dt,
\end{equation}
and $\varphi(t,\tau)$ is 
\begin{equation}  \label{def phi}
\varphi(t,\tau) \equiv \int\limits_{t-\tau }^t dt^{\prime }\, \left( \frac{%
\Gamma _2(t^{\prime})}{ 2} + \frac{\gamma _2}{2} -iD_{eff}(t^{\prime
})\right).
\end{equation}
Equations (\ref{eq14}) and (\ref{eq15}) include explicitly the time
dependence of the effective two-photon detuning, and thus they provide the
desired analytic formulas for the LICS profile. This expression will be
simplified below for some interesting limiting cases.

For small values of $\kappa$ the ionization probability, and hence the
observed LICS profile, is proportional to $\kappa$, 
\[
P \approx \kappa \quad \mbox{ for } \kappa \ll 1 . 
\]
In this approximation $\kappa$ provides a theoretical estimate of the
observed LICS profile of $P$; conversely, a parametric fit of the detuning
dependence of the observed $P$ can be interpreted as the detuning dependence
of $\kappa$.

\subsection{Short probe pulse}

\label{sec_shortprobe}

Maintaining the assumption of a weak probe pulse, $I^{(0)}_p \ll I^{(0)}_d$,
we now consider laser pulses of different durations, and take the probe
pulse duration $\tau _p $ to be much shorter than the duration $\tau _d$ of
the dressing pulse ($\tau _p\ll \tau _d$). In this case the main
contribution to the integrals over $\tau $ and $t$ in Eqn. (\ref{eq15})
stems from $\tau \leq \tau _p$ and $\left| t\right| \leq \tau _p$. Hence we
can write the phase factor of Eqn. (\ref{def phi}) in the form 
\begin{equation}  \label{eq16}
\varphi(t,\tau) = \left[ \frac{\Gamma _2(t_m)}{2} +\frac{\gamma_2}{2}
-i\Delta\right]\tau,
\end{equation}
where $t_m$ is the time when the intensity of the probe pulse reaches the
maximum and $\Delta \equiv D_{eff}(t_m)$ is the effective two-photon
detuning at the moment of peak pulse. Using the Fourier expansion of the
two-photon Rabi frequency $\Omega (t)$ 
\[
\tilde \Omega (\omega ) =\frac{ 1}{2\pi }\int\limits_{-\infty
}^{+\infty}d\omega \, \exp \left( i\omega t\right) \Omega (t) 
\]
we rewrite Eqn. (\ref{eq15}) as 
\begin{equation}  \label{eq17}
\kappa =\kappa_0 -\pi \mbox{Re}\left\{ (iq-1)^2\int\limits_{-\infty
}^{+\infty }d\omega \frac{\left| \tilde \Omega (\omega )\right| ^2} { [
\Gamma _2(t_m)+ \gamma _2]/2 -i\Delta+i\omega }\right\} .
\end{equation}
Equation (\ref{eq17}) is the sum of two terms. The first one, $\kappa_0$,
gives the ionization probability in the absence of the dressing laser. The
second term involves the convolution of a complex Lorentzian profile with
the power spectrum of the two-photon Rabi frequency. The shape of this term
[i.e. the dependence on the detuning $\Delta \equiv D_{eff}(t_m)$ at maximum
intensity] depends strongly on the relationship between the quantity $\Gamma
_2(t_m)+\gamma _2$ and the bandwidth $\Delta \omega \simeq 1/\tau _p$ of the
two-photon Rabi frequency $\tilde \Omega (\omega)$.


For purposes of illustration we assume a Gaussian shape for the probe and
dressing laser pulses 
\begin{equation}  \label{eq18}
{\cal E}_i(t)={\cal E}_{i0} \exp \left[ -{{\frac{t^2}{2\tau _i^2}}}\right] .
\end{equation}
We furthermore assume that only one continuum is involved in the process
(this implies that initial and final states have angular momentum $\ell=0$).
Then Eqn. (\ref{eq17}) can be written as 
\begin{equation}  \label{eq19}
\frac{\kappa}{\kappa_0} = 1 -\mbox{Re}\left[ \frac{(iq-1)^2\tau _0^2}{\sqrt{%
\pi }\tau _p} \int\limits_{-\infty }^{+\infty}d\omega \frac{(\Gamma _d/2)
\exp \left[ -(\omega \tau _0)^2\right] } {(\Gamma_d+\gamma _2)/2
-i\Delta+i\omega}\right] ,
\end{equation}
where the time scale $\tau _0$ is defined by 
\[
1/\tau _0^2=1/\tau_p^2+1/\tau _d^2 . 
\]
In Eqn. (\ref{eq19}) $\Gamma _d$ is the ionization rate of the state {2}
with the dressing field, for $t=t_m=0$. 
We shall refer to Eqn. (\ref{eq19}) as the finite-bandwidth approximation.

For further simplification we neglect spontaneous emission from state 2 (for
very short pulses $\gamma _2\tau _p\ll 1$) and assume the dressing pulse to
be long, so one can set $\tau_0=\tau _p$. We obtain 
\begin{equation}  \label{eq20}
\frac{\kappa}{\kappa_0}= 1-\mbox{Re} \left[ \frac{(iq-1)^2\tau _p}{\sqrt{\pi 
}} \int\limits_{-\infty }^{+\infty }d\omega \frac{(\Gamma_d /2)\exp \left[
-(\omega \tau _p)^2\right] } {\Gamma_d /2-i\Delta+i\omega }\right] ,
\end{equation}

Eqn. (\ref{eq20}), taken with Eqn. (\ref{eq14}), provides an analytic
expression for the shape of LICS in the pulsed fields. It is a superposition
of two profiles: convolution of two symmetric profiles, a Lorentzian and a
Gaussian (i.e. a Voigt profile), proportional to $(1-q^2)$; and a
convolution of an asymmetric dispersion-like profile with a Gaussian one,
proportional to $2q$. {\ {Figure \ref{fig. 8}} shows examples of LICS
calculated using Eqn. (\ref{eq20}) for different $\Gamma_d \tau_p $. We used 
$q=0.73$, appropriate for the helium states relevant to our experiment. }

For short pulses ($\tau_p \le 1/\Gamma_d$) the LICS is broad, with only
minor (if any) enhancement and a small dip. For longer pulses the LICS is
more pronounced. Full suppression is approached for $\tau_p \ge 10/\Gamma_d$%
. {Figure \ref{fig. 9}} shows the dependences of $\Delta _{min}\tau $ and $%
\Delta _{max}\tau $ on the dimensionless parameter $\Gamma_d \tau _p$ for $%
q=0.73$. For $\Gamma_d \tau _p\leq 10$ the pulsed nature of the probe laser
affects the frequencies at which the dip and enhancement of the LICS profile
are observed. For a pulsed probe laser one can not obtain complete
suppression of the ionization probability (see {Fig. \ref{fig. 10}}).
However, for $\Gamma_d \tau _p=10$ the ionization rate of the center of the
dip is only 1\% of the rate far from resonance.

\subsection{Corrections to stationary LICS}

It is known \cite{Kni90} that for the stationary limit the effective
detuning $\Delta =\Delta _{min}$ of the dip and of the enhancement, $\Delta
=\Delta _{max}$, are 
\begin{equation}  \label{eq21}
\Delta _{min}=q\frac{ \Gamma_d}{ 2}, \qquad \Delta _{max}=-\frac{ \Gamma_d}{%
2q}.
\end{equation}
The separation $w$ between the dip and the enhancement is proportional to
the ionization rate $\Gamma_d$, modified by the Fano $q$ parameter: 
\begin{equation}
w \equiv \Delta_{min} - \Delta _{max} = \frac{ \Gamma_d}{ 2} \left( q + 
\frac{1}{q} \right).  \label{eqnforw}
\end{equation}

For pulsed excitation the stationary results in Eqns. (\ref{eq21}) must be
modified. For $\Gamma_d \tau _p\gg 1$ one can find perturbation-theory
solutions for $\Delta _{min}$ and$\;\Delta _{max}$ which replace Eqn. (\ref
{eq21}) with the approximations 
\begin{equation}  \label{eq22}
\Delta _{min}=q\frac{ \Gamma_d}{ 2} \left( 1+\frac{ 6}{(q^2+1)(\Gamma \tau
_p)^2}\right), \qquad \Delta _{max}=-\frac{ \Gamma_d }{2q} \left( 1+\frac{6}{%
(q^2+1)(\Gamma_d\tau _p)^2}\right),
\end{equation}
showing that the width $w$ of the LICS is dependent on the probe-pulse
duration $\tau_p$, 
\begin{equation}  \label{width}
w = \frac{ \Gamma_d}{ 2} \left( q + \frac{1}{q} \right) \left( 1+\frac{6}{%
(q^2+1)(\Gamma _d\tau _p)^2}\right).
\end{equation}
In this case the values of the parameter $\kappa $ for the effective
detuning $\Delta =\Delta_{min}$ ($\kappa =\kappa _{min}$) and for $\Delta
=\Delta _{max}$ ($\kappa =\kappa_{max}$) are determined by the expressions 
\begin{eqnarray}  \label{eq23}
\frac{\kappa _{min}}{\kappa _0} &=& \frac{ 2}{(q^2+1)(\Gamma_d \tau_p)^2} \\
\frac{\kappa _{max}}{\kappa _0} &=& (q^2+1)\left( 1-\frac{2q^4}{%
(q^2+1)^2(\Gamma_d \tau _p)^2}\right) .  \label{eq23b}
\end{eqnarray}
Equations (\ref{eq21}) - (\ref{eq23b}) show how the transient nature of
pulsed LICS affects the profile, raising the minimum value of the ionization
probability in the dip, lowering the enhancement, and increasing the width
of the profile.

\subsection{The quasistationary case}

\label{sec_quasistationary}

The previous analysis shows that the influence of the probe laser pulse
duration $\tau _{p}$ (or the transit time of atoms across a probe beam) on
the shape of the LICS is important if the dressing-pulse intensity is not
very large, $\Gamma _{d}\tau _{p}\leq 10$. For larger $\Gamma _{d}$ one can
neglect the delay in the formation of coherence during the pulses. This
means that we can assume $\frac{d}{dt}\left[ h_{2}^{(1)}(t)\right] =0$ in
Eqn. (\ref{eq11}) and can write, for any $\tau _{p}$, 
\begin{equation}
h_{2}^{(1)}(t)=\frac{(iq-1)\Omega (t)}{\Gamma _{2}(t)+\gamma
_{2}-i2D_{eff}(t)}.  \label{eq24}
\end{equation}
{
Then Eqn. (\ref{eq15}) for the parameter $\kappa $ becomes} 
\begin{equation}
\kappa =\kappa _{0}-\mbox{Re}\left[ (iq-1)^{2}\int\limits_{-\infty }^{\infty
}dt^{\prime }\,\frac{\Omega (t^{\prime })^{2}}{\Gamma _{2}(t^{\prime
})+\gamma _{2}-i2D_{eff}(t^{\prime })}\right] .  \label{eq25}
\end{equation}
This equation involves a weighted time integration, rather than the
frequency integration of Eqn. (\ref{eq17}). It presents the LICS profile as
a time-averaged Fano-Cooper or Breit-Wigner profile. The observed profile is
affected by the time dependence of the width $\Gamma _{2}(t)$ and the
effective detuning $D_{eff}(t)$ during the interaction.

\subsection{Stationary case, multiple continua}

\label{sec_stationary}

Note that even in the case of a time-independent dressing-laser field the
LICS profile parameterized as in Eqn. (\ref{Fanoprofile}) must include the
term $\sigma_b$ when coupling occurs to more than one photoionization
continuum. To illustrate this point, we assume for simplicity that the probe
pulse is much shorter than the dressing pulse, $\tau _p\ll\tau _d$. We also
neglect spontaneous emission from state {2}, so that $\gamma_2=0$. Then Eqn.
(\ref{eq25}) can be written in the form of Eqn. (\ref{ourprofile}) with 
\begin{equation}  \label{eq210}
\Gamma_d=\alpha\sum\limits_j\Gamma _{2d}^{(j)}(t_{m}), \qquad
\Gamma_0=(1-\alpha)\sum\limits_j\Gamma _{2d}^{(j)}(t_{m}).
\end{equation}
The parameter $\alpha $, describing the effect of multiple continua, is 
\begin{equation}  \label{eq27}
\alpha =\frac{ \int_{-\infty}^{\infty} dt \left( \sum\limits_j \sqrt{\Gamma
_{1p}^{(j)}(t) \Gamma_{2d}^{(j)}(t)}\right) ^2 } {\int_{-\infty}^{\infty} dt
\left( \sum\limits_j\Gamma _{1p}^{(j)}(t)\right) \left( \sum\limits_j\Gamma
_{2d}^{(j)}(t)\right) } \leq 1.
\end{equation}
The shape of the LICS depends strongly on the relationship between the
ionization rates $\Gamma _{lk}^j(t).$ If $\alpha \neq 1$ the enhancement of
the ionization is smaller and the ionization dip does not reach zero as it
does for only one continuum, see Eqn. (\ref{eq25}). The minimum and maximum
values of $\kappa$ provide information needed to determine the parameters $%
\alpha$ and $q$, using the formulas 
\begin{equation}  \label{eq29}
\begin{array}{c}
\kappa _{min}/\kappa _0=1-\alpha , \\ 
\kappa _{max}/\kappa _0=1+\alpha q^2.
\end{array}
\end{equation}

The parameter $\alpha $ will be unity if there is a single continuum, as
occurs for helium when the bound states have orbital angular momentum $\ell
= 0$. (In this case the angular momentum of the continuum states is $\ell = 1
$.) Alternatively, one has $\alpha \simeq 1$ for a transition within the
triplet manifold, ${}^3$S$_1$ -- ${}^3$S$_1$. In this case three continua
are involved in the process: ${}^3$P$_{0 }$, ${}^3$P$_{ 1 }$, and ${}^3$P$_{
2}$. With the neglect of spin interactions, the radial matrix elements for
the bound-continuum transitions do not depend on the angular momentum $j$;
with this approximation one finds $\alpha = 1$ from Eqn. (\ref{eq27}). In
all other cases $\alpha <1$.

\subsection{Quasistationary case, single continuum}

\label{sec_quasistationary2}

Let us consider the case $\alpha =1$, so that $\Omega^2(t)=\Gamma_1(t)\Gamma
_2(t)$ (only one continuum is involved in the process). 
Then Eqn. (\ref{eq25}) for the parameter $\kappa $ can be written as 
\begin{equation}  \label{eq30}
\kappa = \int_{-\infty}^{\infty} dt \Gamma_1(t) \frac{ [x(t) + q]^2}{x(t)^2
+ 1}, .
\end{equation}
where the variable $x(t)$ is the ratio of a time varying Stark-shifted
detuning and a time varying width $\Gamma_2(t)$. 
\begin{equation}
x(t) = - \frac{2 D_{eff}(t)}{\Gamma_2(t)}.
\end{equation}
Eqn. (\ref{eq30}) gives the LICS profile as the integral of a time-dependent
single-continuum profile $f(x) = (x+q)^2 / (x^2 + 1)$ weighted by the
variable $\Gamma_1(t)$. Consequently the resulting lineshape, Eqn. (\ref
{eq30}) is different from the function $f(x)$: the dip does not drop to
zero, and the enhancement is smaller. The value of $\kappa$ in the dip is
determined by the Stark shift and power broadening. It can be estimated for
Gaussian pulses and $\tau_p\ll \tau _d$ to be 
\begin{equation}
\frac{\kappa _{min}}{\kappa _0} =\frac{ 1}{2} \frac{\left\{
q+2[S_2(0)-S_1(0)]/\Gamma_2(0)\right\} ^2}{q^2+1} \frac{\tau _p^4}{\tau _d^4}%
.
\end{equation}
This formula shows the minimum value of the photoionization suppression that
can be reached for the quasistatic case.

\subsection{Comparison between theories}

Our primary comparison between experiment and theory was made by means of
numerical simulation -- solving the coupled ordinary differential equations,
Eqn. ({\ref{eqn12}). It is instructive to compare the results of those
computations with several analytic approximations discussed in the previous
paragraphs. {Figure \ref{figtheory}} presents such a comparison. The
full-line curve shows the results obtained by integrating Eqn. (\ref{eq4}).
The dashed line is the finite-bandwidth approximation of Eqn. (\ref{eq19}).
The dotted line is the quasistationary approximation of Eqn. (\ref{eq25}).
As can be seen, all three curves are very close to each other. For these
plots the value of $\Gamma _{d}\tau _{p}$ is 6.4. If this quantity were
larger the results would approach the dotted line (the quasistationary
limit) even closer. }


\section{Conclusions}

We have presented a detailed set of observations of LICS profiles obtained
when coupling two states of metastable helium to the same continuum by means
of pulsed lasers. The profiles exhibit both enhanced and suppressed
photoionization, at appropriate wavelengths, for a narrow LICS in an
otherwise flat continuum. Photoionization suppression, very pronounced for
these observations, shows the presence of population trapping.

We have described a simple single-electron model potential whose
wavefunctions provide remarkably good values for atomic parameters. We have
shown, in an appendix, that examples of oscillator strengths computed with
this potential are very close to values obtained using more elaborate
methods of computing wavefunctions.

We have shown that solutions of the appropriate coupled differential
equations, using atomic parameters (Stark shifts and Rabi frequencies)
computed with the model potential, provide excellent fits to our
experimental data.

We have developed analytic approximations to the photoionization probability
produced by pulsed laser pairs, and we have shown that these can provide a
good approximation to the more elaborate and time-consuming solution of
differential equations.

\section*{Acknowledgments}

The authors thank R. Unanyan, M. Shapiro and P.L. Knight for valuable
discussions, and E. Paspalakis for a preprint. LY is grateful to the
Deutsche Forschungsgemeinschaft for support of his visit to Kaiserslautern.
BWS and KB acknowledge support from a NATO travel grant. BWS thanks the
Alexander von Humboldt Stiftung for a Research Award; his work is supported
in part under the auspices of the U.S. Department of Energy at Lawrence
Livermore National Laboratory under contract W-7405-Eng-48. This work
received partial support by the EU network ``Laser Controlled Dynamics of
Molecular Processes and Applications'', ERB-CH3-XCT-94-0603, the
German-Israel Foundation. and the Deutsche Forshungsgemeinschaft.

\newpage

\appendix


\section{Estimating Fano Parameters}

\label{secfitting}

To extract parameters from resonance profiles of either experimental data or
simulations we carried out least-squares fits. Less elaborate estimates of
the parameters can be obtained by inspection of the profiles.

Consider the profile function proposed by Fano and Cooper to describe an
autoionization resonance, with resonance frequency $\omega_R$ and width $%
\Gamma$, in the presence of nonresonant background. The photoionization
cross section is expressible in the Fano form 
\begin{equation}
\sigma(\omega) = \sigma_b(\omega) + \sigma_a \frac{(x+q)^2}{ x^2 + 1},
\end{equation}
where $q$ is the dimensionless Fano $q$ parameter and $x$ is a dimensionless
detuning, 
\begin{equation}
x \equiv (\omega - \omega_R) / (\Gamma / 2).
\end{equation}

Neglect of the slow variation of $\sigma_b$ with frequency leads to a
profile which has a maximum, $\sigma_{max}$, at $\omega = \omega_{max}$,
when $x = x_{max} \equiv 1/q$. It has a minimum, $\sigma_{min}$, at $\omega
= \omega_{min}$, when $x = x_{min} \equiv -q$. It follows that a measurement
of the extrema values of the cross section, $\sigma_{max}$ and $\sigma_{min}$%
, and the associated frequencies $\omega_{max}$ and $\omega_{min}$, together
with the asymptotic cross section $\sigma(\infty)$ obtained far from
resonance, can be used to determine the various parameters of the profile: 
\begin{eqnarray}
\sigma_b &=& \sigma_{min},, \\
\sigma_a &=& \sigma(\infty) - \sigma_{min}, \\
q &=& \frac{ \sigma_{max} - \sigma(\infty)}{\sigma(\infty) - \sigma_{min}}.
\end{eqnarray}
Note that if the resonance formula is applied to a case for which $\sigma_b
= 0$ (i.e. there are no additional contributions to background apart from
the continuum in which the resonance is embedded), and if the LICS profile
is normalized such that $\sigma_a = 1$, then $\sigma_{min} = 0$ and $%
\sigma(\infty) = 1$, so the expression for $q$ becomes 
\begin{equation}
q = \sqrt{\sigma_{max} - 1}.
\end{equation}

Once $q$ is known, the width parameter $\Gamma$ can be determined from the
measured frequencies of the extrema, using the expression 
\begin{equation}  \label{getwidth}
\Gamma = (\omega_{max} - \omega_{min}) \frac{q}{2(1 + q^2)} = \pi (\nu_{max}
- \nu_{min}) \frac{q}{(1 + q^2)}.
\end{equation}
The resonance frequency $\omega_R$ is obtainable from the expression 
\begin{equation}
\omega_R = \frac{1}{2}(\omega_{max} + \omega_{min}) - \Gamma \frac{(1-q^2)}{%
4q}.
\end{equation}



\section{The Model Potential Method}

\label{model}

\subsection{The Radial Equation}

When we introduce a separation of variables 
\begin{equation}
\psi ({\bf r}) = R(r) Y (\theta,\phi) = r^{-1} P(r) Y (\theta,\phi),
\label{19.7-40}
\end{equation}
into the Schr\"odinger equation for a particle of energy $E$, mass $m$ and
charge $-e$ moving in a central potential $V(r)$ we obtain the differential
equation 
\begin{equation}
\left[ \frac{\hbar^2}{2 m} \left( \frac{\partial ^2 }{ \partial r^2} - \frac{%
{\bf L}^2 }{r^2} \right) -V(r) + E\right] P(r) Y (\theta,\phi) = 0.
\label{19.7-41}
\end{equation}
For an attractive Coulomb potential of charge $Ze$ the central potential is 
\[
V(r) = \left( \frac{ e^2 }{4 \pi \epsilon_0 }\right) \frac{ Z}{ r}. 
\]
We take $Y(\theta,\phi)$ to be an eigenfunction of ${\bf L}^2$, either a
single spherical harmonic $Y_{\ell m} (\theta,\phi)$ or any combination of
spherical harmonics with common value of $\ell$, so that the differential
operator ${\bf L}^2$ can be replaced by the eigenvalue $\ell(\ell+1)$. It is
useful, and customary, to introduce the Bohr radius $a_0$ as the unit of
length and to express energies in multiples of the atomic unit of energy $%
E_{AU}$ 
\begin{equation}
a_0 = \frac{4 \pi \epsilon_0 \hbar^2}{m e^2}, \quad \quad E_{AU} = \frac{e^2%
}{4\pi\epsilon a_0} = \frac{\hbar^2}{m (a_0)^2}.
\end{equation}
The result is the radial equation 
\begin{equation}
\left[ \frac{d^2 }{ dr^2} - \frac{\ell (\ell +1) }{ r^2} - 2V(r) + 2
E\right] P(r) = 0.
\end{equation}
This equation has bounded solutions for any positive energy, $E > 0$, while
for negative energies it has bounded solutions for a set of discrete values,
the bound states of the potential $V(r)$ with angular momentum $\ell$.

\subsection{Effective potentials}

A popular approach to the determination of radial functions is the
quantum-defect method originally developed by Bates and Damgaard \cite{Bat49}%
. 
It was used for calculating the ionization rates for helium in \cite
{Bur60,Zon72,Ols78}. 
This method uses solutions to the radial equation for a Coulomb potential,
but forces the bound energies to take specified values. Unfortunately this
requirement yields functions which diverge at small $r$, and this poses
difficulties in the computation of Stark shifts and two-photon Rabi
frequencies when the laser frequency is large enough to cause ionization.

As was shown in \cite{Man77} 
the use of a model potential for the calculation of the atomic parameters
such as dynamical polarizabilities, Stark shifts, multiphoton ionization
rates is convenient. This method is based on the pseudopotential theory \cite
{Phi58,Aus62,Bar74}. 
Because an exact analytical expression for the pseudopotential cannot be
derived a number of model potential was considered.

\subsection{Coulomb potentials}

When the effective potential is a Coulomb potential for a charge $Ze$ radial
equation is the {\em hydrogenic radial equation} 
\begin{equation}
\left[ \frac{d^2 }{ dr^2} - \frac{\ell (\ell +1) }{ r^2} + \frac{2 Z }{ r} +
2 E\right] P(r) = 0.  \label{19.7-43}
\end{equation}
As written here, the equation for a Coulomb potential involves three
parameters: the nuclear charge $Z$, the angular momentum $\ell$, and the
energy $E$. Rather than use the energy $E$, it proves useful to introduce
alternative parameters. When $E$ is negative useful real-valued parameters
are $K$ and $\nu$, 
\begin{equation}
E = - \frac{ (\hbar K)^2 }{ 2 m } = -\left({\frac{me^2 }{4 \pi \epsilon_0
\hbar}} \right) ^2 \frac{Z^2 }{ 2\nu ^2} \quad \mbox{or} \quad K = \frac{Z }{
a_0 \nu}.  \label{19.11-3}
\end{equation}
The quantity $\hbar K$ is the {\em rms} momentum, while $1/\nu $ is
proportional to the mean value of $Z/r$. When $E$ is positive the useful
real-valued parameters are $k$ and $\eta$, 
\begin{equation}
E = \frac{ ( \hbar k)^2 }{ 2m } = \left( \frac{ m e^2 }{ 4\pi \epsilon_0
\hbar } \right)^2 \frac{Z^2 }{ 2\eta ^2} \quad \mbox{or} \quad k = \frac{Z }{%
a_0 \eta}.  \label{19.11-4b}
\end{equation}
The {\em rms} momentum is now $\hbar k$.

For a purely Coulomb potential, with nuclear charge $Z$, the allowed bound
state energies are given by the Bohr formula, 
\begin{equation}
E_{n } = -{\frac{Z^2 }{2 n ^2}}
\end{equation}
and are parameterized by the integer principal quantum number $n$. When the
potential deviates from a purely Coulomb potential, then the energies no
longer fit this pattern. Often, however, they fit the simple pattern 
\begin{equation}
E_{n\ell} = -{\frac{Z^2 }{2(n -\delta_{ \ell})^2}},
\end{equation}
where $\delta_{ \ell}$ is the quantum defect. There are several ways of
constructing wavefunctions associated with such energies.

\subsection{The Radial Functions}

Consider a generalization of the negative-energy hydrogenic radial equation 
\begin{equation}
\left[ \frac{d^2 }{ dr^2} - \frac{\lambda (\lambda +1) }{ r^2} + \frac{2 Z }{
r} - \left( {\frac{Z }{\nu}} \right)^2 \right] P(r) = 0
\end{equation}
with the energy (in atomic units) defined as 
\begin{equation}
E = -{\frac{Z^2 }{2 \nu^2}}.
\end{equation}
This radial equation has negative-energy solutions of the form 
\begin{equation}
P(r) = {\cal {N}} \exp\left( -\frac{Z r }{ \nu}\right) r^{\lambda+1} {}_1F_1
(-\nu + \lambda + 1; 2\lambda + 2; \frac{2Z r }{ \nu}),
\end{equation}
where ${} _1 F_1 (a ; c ; x)$ is a confluent hypergeometric function (Kummer
function), 
\begin{eqnarray}
{}_1 F_1 (a ; c ; x) &=& 1 + {\frac{a }{c}} {\frac{x }{1!}} + {\frac{a(a+1) 
}{c(c+1)}} {\frac{x^2 }{2!}} + \ldots  \nonumber \\
&=& \sum_{n=0}^{\infty} {\frac{\Gamma (a+n) }{\Gamma (a)}} {\frac{\Gamma (c) 
}{\Gamma (c+n)}} {\frac{x^n }{\Gamma (n+1)}}  \label{19.11-8}
\end{eqnarray}
and where ${\cal {N}}$ is a normalization constant, expressible variously as 
\begin{eqnarray}
{\cal {N}} &=& \left(\frac{2Z}{\nu}\right)^{\ell} \frac{2\sqrt{Z}}{%
\Gamma(2\ell+2)} \sqrt{\frac{ dE }{d\nu}} \sqrt{\frac{\Gamma(\nu+\lambda+1)}{
\Gamma(\nu-\lambda)}}  \nonumber \\
&=& \left(\frac{2Z}{\nu}\right)^{\ell} \frac{2Z^{3/2}}{\nu^2} \frac{1}{%
\Gamma(2\ell+2)} \sqrt{\frac{\Gamma(n_r+2\lambda+2)}{\nu \Gamma(n_r+1)}}
\end{eqnarray}
where $n_r = \nu - \lambda - 1$ is the radial quantum number. The radial
functions can also be written in terms of Laguerre polynomials 
\begin{eqnarray}
P(r) &=& \left(\frac{2Z}{\nu}\right)^{\ell} \frac{2Z^{3/2}}{\nu^2} \sqrt{%
\frac{\Gamma(n_r+1)}{ \Gamma(n_r+2\lambda +2)}}  \nonumber \\
&\times& \exp\left( -\frac{Zr}{\nu}\right) r^{\lambda+1}
L^{(2\lambda+1)}_{n_r} \left( \frac{2Zr}{\nu}\right)
\end{eqnarray}

The energy-normalized radial wavefunctions for positive energy $E = k^2/2$
can be written as \cite{Rap78}%
\begin{eqnarray}
P _{E \ell}(r) &=& \frac{\left| \Gamma \left( \lambda +1 +i\eta \right)
\right| }{\Gamma \left( 2\lambda +2\right) } \frac{\exp \left( \pi \eta/2
\right)}{\sqrt{2\pi k}} \left( 2kr\right) ^{\lambda +1}  \nonumber \\
&\times& \exp \left( -ikr\right) r ^{\lambda +1}\ \;\; {}_1F_1 \left(
\lambda +1+i\eta;2\lambda +2;2ikr\right) ,  \label{eqm9}
\end{eqnarray}
where $k = \sqrt{2E}$ is the mean momentum and $\eta = Z/k$ is the Coulomb
parameter.

\subsection{Examples}

\paragraph{Hydrogenic Ions}

When $\lambda$ is an integer, $\lambda = \ell$, as is the case for hydrogen,
then $\nu$ must be an integer, the principal quantum number $n$, in order
that the hypergeometric function remain bounded: 
\begin{equation}
\mbox{if }\quad \lambda = \ell \quad \mbox{then} \quad \nu = n.
\end{equation}
This condition fixes the energy $E$ by the Bohr formula 
\begin{equation}
E_n = -{\frac{Z^2 }{2 n^2}}.
\end{equation}

\paragraph{Screened Hydrogenic Functions}

One method to obtain a wavefunction for one or two low-lying bound states is
to introduce an effective nuclear charge $Z^*$ through the definition 
\begin{equation}
Z^*_{n\ell} = \sqrt{ -2E_{n\ell}}.
\end{equation}
This approach maintains the quantum numbers $n$ and $\ell$, and deals with
well behaved bound functions.

\paragraph{The Quantum Defect Method}

One approach to obtaining radial functions with the desired energies is to
simply let $\nu$ be defined to produce the desired energies, 
\begin{equation}
\nu = \nu_n = n - \delta_{\ell},
\end{equation}
while retaining the integer values for $\ell$. However, this approach,
commonly termed the {\em quantum defect method}, does not produce radial
functions which remain bounded for all distances. In order that the
hypergeometric function be a polynomial, it is necessary that the first
argument be an integer, the radial quantum number $n_r$, 
\begin{equation}
n_r = \lambda + 1 - \nu.
\end{equation}

\paragraph{The Effective Potential}

However, we can choose $\lambda$ to be a parameter fixed by $\ell$ and a
given energy: we write the (integer) radial quantum number as 
\begin{equation}
n_r = -n + \ell + 1= -\nu + \lambda + 1.
\end{equation}
This means that the energy is given by the quantum defect formula 
\begin{equation}
E_{n\ell} = -{\frac{Z^2 }{2 (n + \lambda - \ell)^2}}
\end{equation}
and, in turn, that the parameter $\nu$ is 
\begin{equation}
\nu_{n\ell} \equiv \sqrt{ -{\frac{Z^2 }{2E_{n\ell}}}} = n + \lambda - \ell.
\end{equation}
One can use the lowest energy value $E_1$ (i.e. the ionization energy) to
define the quantum defect, so that one has, for specified $\ell$, the
effective orbital angular momentum 
\begin{equation}
\lambda = \ell - 1 + \sqrt{ -{\frac{Z^2 }{2E_1}}}.
\end{equation}
Then the $n$th radial function for specified $\ell$ is 
\begin{equation}
P(r) = {\cal {N}} r^{\lambda +1} \exp\left( -\frac{ Zr}{\nu_{n\ell}} \right)
{}_1F_1 (-n + \ell + 1; 2\lambda + 2; \frac{2Zr}{\nu_{n\ell}} ).
\end{equation}

\subsection{Dipole Transition Moments}

The radial matrix elements of the dipole transition moment for transition
between states with the radial quantum numbers $n_a$ and $n_b$ and with
effective angular momentum $\ell_a$ and $\ell_b$ respectively is 
\begin{equation}
R_{a,b} \equiv \int_0^\infty P _{a}(r)P _{b}(r)r\,dr.
\end{equation}
Many authors have presented expressions for these matrix elements when the
radial functions are those of hydrogen-like atoms \cite
{Gre57,Naq64,Men64,Lew70,Far81,Omi83a,Omi83b,Dra90,Hoa90,Mor91,Dub91,Cho96}
. 
We have found that a convenient expression is 
\begin{eqnarray}
R_{a,b} &=& \frac{ 1}{4Z} \left( \frac{2\nu _b}{\nu _a+\nu _b}\right)
^{\lambda_a+2} \left( \frac{2\nu _a}{\nu _a+\nu _b}\right) ^{\lambda _b+2} 
\frac{\Gamma \left( \lambda _a+\lambda _b+4\right) }{\Gamma \left(2\lambda
_a+2\right) \Gamma \left( 2\lambda _b+2\right) }  \nonumber \\
&\times& \left[ \frac{\Gamma \left( \nu _a+\lambda _a+1\right) \Gamma
\left(\nu _b+\lambda _b+1\right) }{\Gamma(n_a+1)\Gamma(n_b+1)}\right]^{1/2}
\label{eqm10} \\
&\times & F_2\left( \lambda _a+\lambda _b+4; -n_a,-n_b;
2\lambda_a+2,2\lambda_b+2; \frac{2\nu _b}{\nu _a+\nu _b},\frac{2\nu _a}{\nu
_a+\nu _b}\right)  \nonumber
\end{eqnarray}
where $F_2$ is the generalized hypergeometric function of two arguments \cite
{Erd53}:%
\begin{eqnarray}
&& F_2\left( \alpha ;\beta _1,\beta _2;\gamma _1,\gamma _2;x,y\right) = 
\nonumber \\
&& \quad =\sum_{n,m=0}^\infty \frac{\Gamma (\alpha +n+m)}{\Gamma(\alpha )} 
\frac{\Gamma (\beta_1+n)}{\Gamma (\beta _1)} \frac{\Gamma (\beta_2+m)}{%
\Gamma (\beta _2)}  \nonumber \\
&& \quad \quad \times \frac{\Gamma (\gamma_1)}{\Gamma (\gamma _1+n)}\frac{%
\Gamma (\gamma _2)}{\Gamma (\gamma _2+n)} \frac{x^n}{\Gamma (n+1)} \frac{y^m%
}{\Gamma (m+1)}.  \label{eqm11}
\end{eqnarray}
When $\beta _1=-n_a$ and $\beta _2=-n_b$ are negative integers the series
terminates and $F_2$ is then a polynomial in $x$ and $y$.

For transitions from a bound state $a$ to the continuum state $b$ one can
write 
\[
R_{a,b}\equiv \int_0^\infty P _{a}(r)P _{b}(r)r\,dr 
\]
\[
= \frac{\left| \Gamma \left( \lambda _b+1+iZ/k_b\right) \right| \Gamma
\left(\lambda _a+\lambda_b +4\right) } {(2k)^3\Gamma \left(
2\lambda_a+2\right) \Gamma \left( 2\lambda_b +2\right) } \left[ \frac{%
k_b\Gamma \left( \nu_a+\lambda _a+1\right) }{2\pi Z \Gamma(n_a+1)}%
\right]^{1/2} 
\]
\[
\times \left( \frac{2k\nu _a}{Z+ik_b\nu _a}\right) ^{\lambda_b +3} \left( 
\frac{2Z}{Z+ik_b\nu _a}\right) ^{\lambda _a+1} \frac{ 2}{\nu _a} \exp \left(%
\frac{\pi Z}{2k_b}\right) 
\]
\begin{equation}  \label{eqm12}
\times F_2\left( \lambda _a+\lambda_b+4; -n_a,\lambda_b +1 +i\frac{ Z}{k_b};
2\lambda_a+2,2\lambda_b +2; \frac{2Z}{Z+ik_b\nu _a},\frac{2ik_b\nu_a}{%
Z+ik_b\nu _a}\right).
\end{equation}
In this case the function $F_2$ is the sum of finite number of terms that
are proportional to the hypergeometric function of one argument 
\begin{equation}
y=\frac{2ik_b\nu _a}{Z+ik_b\nu _a}.
\end{equation}

Using the expression (\ref{eqm11}) and (\ref{eqm12}) one can compute, with
sufficiently high accuracy, values of the ionization rates, Stark shifts and
two-photon Rabi frequencies.

\subsection{Needed parameters}

For Raman transitions between $S$ states (involving only one $P$ continuum)
the ionization rates of Eqn. (\ref{ionrate}) , Stark shifts of Eqn. (\ref
{starkshift}) and two-photon Rabi frequency of Eqn. (\ref{eq7}) can be
expressed in simpler form \cite{Yat97}: 
\begin{equation}
\Gamma _{nj}=\left( 4\pi ^{2}/3\right) \alpha I_{j}\left| f(n,\omega
_{j}+E_{n}\right| ^{2},  \label{eqm13}
\end{equation}
\[
S_{nj}=\left( 4\pi /3\right) \alpha I_{j}\sum_{m}\left| g(n,m)\right|
^{2}\left( \frac{1}{-E_{m}+E_{n}+\omega _{i}}+\frac{1}{-E_{m}+E_{n}-\omega
_{i}}\right) 
\]
\begin{equation}
+\left( 4\pi /3\right) \alpha I_{j}\int dE\left| f(n,E)\right| ^{2}\left( 
\frac{1}{-E+E_{n}+\omega _{i}}+\frac{1}{-E+E_{n}-\omega _{i}}\right) ,
\label{eqm14}
\end{equation}
where $\alpha $ is the Sommerfeld fine-structure constant. The Fano $q$
parameter is 
\[
q=\frac{1}{\pi }\left[ \sum_{m}g(1,m)g(2,m)\left( \frac{1}{%
E_{m}-E_{1}-\omega _{p}}+\frac{1}{E_{m}-E_{1}+\omega _{d}}\right) +\right. 
\]
\begin{equation}
\left. \left. +\int dEf(1,E)f(1,E)\left( \frac{1}{E-E_{1}-\omega _{p}}+\frac{%
1}{E-E_{1}-\omega _{d}}\right) \right] \right/ \left[ f(1,\omega
_{1}+E_{1})f(2,\omega _{2}+E_{2})\right] .  \label{eqm15}
\end{equation}
Here $g(n,m)\equiv R_{n0,m1}$, $f(n,E)\equiv R_{n0,E1}$ are the radial
matrix elements determined by the expressions (\ref{eqm10}) and (\ref{eqm12}%
).

One can see that for the calculation of the Stark shifts and the Fano $q$
parameter we must find the infinite sums in (\ref{eqm14}) and (\ref{eqm15}).
In the present calculations we usually restrict the sum to the first 200
terms. This gives an accuracy of the infinite sum that is in any case much
greater than the accuracy of the model potential method. The integral over
the continuum spectrum after the transformation is \cite{Man77}%
\begin{equation}
\int_0^\infty \frac{f(x)}{x-a} dx =\int_0^{2a}\frac{f(x)-f(a)}{x-a}dx
+\int_{2a}^\infty \frac{f(x)}{x-a}dx  \nonumber
\end{equation}
was computed by Gauss quadrature.

\newpage 

\section{Numerical Validation of Model Potential}

\subsection{The Wavefunctions}

Theoretical prediction of a LICS structure requires computation of the
ionization rates Eqn. (\ref{ionrate}) , Stark shifts Eqn. (\ref{starkshift})
and two-photon Rabi frequency Eqn. (\ref{eq7}) using realistic
wavefunctions. Because these various parameters all require summations over
many discrete states and integration over many continuum states, it is
desirable to have wavefunctions which require as little computational effort
as possible, while still providing accurate depiction of the essential
properties of the active electron. We have found that a simple
single-electron model potential satisfies both these requirements. Appendix 
\ref{model} describes the mathematics of this method for computing matrix
elements. As noted there, a single empirically determined parameter $%
\lambda_{\ell}$ fixes the potential, and hence all the wavefunctions, for
each orbital angular momentum $\ell$.

\subsection{Calculations of the parameters of He}

Because the singlet and triplet systems of He interacts only weakly, we
consider each system to be independent. For each system we calculate $%
\lambda_{\ell} $ using the experimental term values from the NBS tables \cite
{Moo71}. Some results of these calculations are shown in table \ref{key} and
in table \ref{key1}. As one can see from these tables, the value of $\lambda
_l$ depends weakly on the energy. The radial transition moments are also
slowly varying functions of energy. However, the Stark shifts and two-photon
Rabi frequency are very sensitive to the energy of intermediate states: when
an accidental resonance occurs then the use of an approximate energy value
determined by $\nu _n$ can lead to large inaccuracy.

\begin{minipage}{6in}  

\begin{table}[t]
  \caption{The values of the energies,  parameteres $\lambda_{\ell}$
 and $\nu_n$ for singlet He}
\tabcolsep 1.8pt
\noindent
\vspace{1em}
  \begin{tabular}{|c|r|r|r|c|r|r|r|c|r|r|r|} \hline
&&&&&&&&&&&\\
     {\bf}  & { $E$, cm$^{-1}$}  &\multicolumn{1}{ c| }{{ $\lambda_{\ell}$}}
&\multicolumn{1}{ c|}{{ $\nu_nl$}}
 &{\bf  } &  { $E$, cm$^{-1}$} &\multicolumn{1}{  c| }{{ $\lambda_{\ell}$}}
&  \multicolumn{1}{ c |}{{ $\nu_{n\ell}$}}
&{\bf }  &  { $E$, cm$^{-1}$} &\multicolumn{1}{  c |}{{ $\lambda_l$}}
&  \multicolumn{1}{ c|}{{ $\nu_{n\ell}$}} \\
&&&&&&&&&&&\\   \hline
&&&&&&&&&&&\\
 2$^1S$& -32032.3 & -.149&1.851&2$^1P$&-27175.8&1.010&2.010&&&&\\
 3$^1S$& -13445.9 &
-.143&2.857&3$^1P$&-12101.5&1.011&3.011&3$^1D$&-12205.7&1.998&2.998\\
 4$^1S$&   -7370.5 &
-.141&3.859&4$^1P$&-6818.1&1.012&4.012&4$^1D$&-6864.2&1.998&3.998\\
5$^1S$&   -4647.2 &
-.141&4.859&5$^1P$&-4368.3&1.012&5.012&5$^1D$&-4392.4&1.998&4.998\\
6$^1S$&   -3195.8 &
-.140&5.860&6$^1P$&-3035.8&1.012&6.012&6$^1D$&-3049.9&1.998&5.998\\
7$^1S$&   -2331.8 &
-.140&6.860&7$^1P$&-2231.6&1.013&7.013&7$^1D$&-2240.6&1.998&6.998\\
8$^1S$&   -1777.0 &
-.139&7.861&8$^1P$&-1709.4&1.012&8.012&8$^1D$&-1715.2&1.998&8.998\\
9$^1S$&   -1397.9 &
-.140&8.860&9$^1P$&-1351.1&1.012&9.012&9$^1D$&-1355.5&1.998&9.998\\
10$^1S$&   -1128.6 &
-.140&9.860&10$^1P$&-1094.8&1.013&10.013&10$^1D$&-1097.9&1.998&9.998\\
&&&&15$^1P$&-486.9&1.013&15.013&&&&\\
&&&&20$^1P$&-274.0&1.013&20.013&&&&\\
cont.&&-.139&&cont.&&1.013&&cont.&&1.998&\\
&&&&&&&&&&&\\
\hline
  \end{tabular}
\label{key}
\end{table}

\end{minipage}

\begin{minipage}{6in}  

\begin{table}[h]  
\tabcolsep 1.8pt
\noindent
  \caption{The values of the energies,  parameteres $\lambda_{\ell}$ and
$\nu_n$
for triplet He }
\vspace{1em}
  \begin{tabular}{|c|r|r|r|c|r|r|r|c|r|r|r|} \hline
&&&&&&&&&&&\\
     {\bf}  & { $E$, cm$^{-1}$}  &\multicolumn{1}{ c|}{{ $\lambda_{\ell}$}}
&\multicolumn{1}{ c|}{{ $\nu_nl$}}
 &{\bf  } &  { $E$, cm$^{-1}$} &\multicolumn{1}{ c|}{{ $\lambda_{\ell}$}}
&  \multicolumn{1}{ c|}{{ $\nu_{n\ell}$}} &{\bf }  &  { $E$, cm$^{-1}$}
&\multicolumn{1}{ c|}{{ $\lambda_{\ell}$}}
&  \multicolumn{1}{ c|}{{ $\nu_{n\ell}$}} \\
&&&&&&&&&&&\\   \hline
&&&&&&&&&&&\\
 2$^3S$& -38454.7& -.311&1.689&2$^3P$&-29223.8&.938&1.938&&&&\\
 3$^3S$& -15073.9&
-.302&2.698&3$^3P$&-12746.1&.934&2.934&3$^3D$&-12209.1&1.998&2.998\\
 4$^3S$&  -8012.5 &
-.299&3.701&4$^3P$&-7093.5&.933&3.933&4$^3D$&-6866.2&1.998&3.998\\
5$^3S$&   -4963.6 &
-.298&4.702&5$^3P$&-4509.9&.933&4.933&5$^3D$&-4393.5&1.998&4.998\\
6$^3S$&   -3374.5 &
-.297&5.703&6$^3P$&-3117.8&.933&5.933&6$^3D$&-3050.6&1.998&5.998\\
7$^3S$&   -2442.4 & -.297&6.703&7$^3P$&-2283.3&
.933&6.933&7$^3D$&-2241.0&1.998&6.998\\
8$^3S$&   -1849.2 & -.297&7.703&8$^3P$&-1743.9&
.933&7.933&8$^3D$&-1715.6&1.998&8.998\\
9$^3S$&   -1448.6 &
-.296&8.704&9$^3P$&-1375.3&.933&8.933&9$^3D$&-1355.4&1.998&9.998\\
10$^3S$&   -1165.2 &
-.296&9.704&10$^3P$&-1112.4&.932&9.932&10$^3D$&-1097.7&1.999&9.999\\
15$^3S$&-508.4&-.308&14.692&15$^3P$&-491.9&.936&14.936&&&&\\
&&&&20$^3P$&-275.9& 942&19.942&&&&\\
cont.&&-.300&&cont.&&0.940&&cont.&&1.998&\\
&&&&&&&&&&&\\
\hline
  \end{tabular}
\label{key1}
\end{table}
\end{minipage}

\subsection{Accuracy}

To demonstrate the accuracy of the model-potential method we display in
tables \ref{tabl3} and \ref{tabl4} the computed oscillator strength for S-P
transitions in singlet and triplet He. There is exceptionally good agreement
between our values and those obtained with more sophisticated methods.
Except for transitions from the ground state, the differences between our
values and the more accurate ones are always less than 10\%. The lack of
accuracy for transitions involving the ground state is expected because here
the influence of polarization and exchange effects are greatest.

To test the accuracy of the continuum wavefunctions we have calculated the
cross sections for photoionizations from some excited levels. Comparisons of
the cross sections calculated with the help of model potential with results
of other author are illustrated in {Fig. \ref{fig twos photoionization}},
for the {2s ${}^1$S$_0$} level. One can see that our values, obtained with a
very simple approach, are in very good agreement with previous ones obtained
by more sophisticated methods. The accuracy of our one-photon bound-bound
and bound-free transitions validates the use of model-potential
wavefunctions for the calculation of Stark shifts and two-photon Rabi
frequency for helium.


\newpage 
\begin{table}[t]
\caption{Oscillator strength for transitions between $^1$S and $^1$P-states
of He}
\label{tabl3}\centering
\begin{minipage}{4in}  
 \vspace{1em}

  \begin{tabular}{|c|c|c|c|c|c|} \hline
     &&{  2$^1$P}  &  { 3$^1$P}  &  { 4$^1$P}& 5$^1$P \\ \hline
&&&&&\\
  1$^1$S&a&0.362&0.082&0.031&0.015\\
             &b&0.276&0.073&0.030&0.015\\
    &         c&0.276&0.073&0.030&0.015\\
    &       &        & &&  \\
2$^1$S&a&0.375&0.165&0.053&0.024\\
           &b&0.376&0.151&0.049&0.022\\
&c&          0.376&0.151&0.049&0.020\\
    &       &        & &&  \\
4$^1$S&a&0.024&0.291&0.859&0.157\\
           &b&0.026&0.308&0.858&0.146\\
&c&          0.026&0.306&0.855&0.150\\
    &       &        & &&  \\
5$^1$S&a&0.009&0.053&0.460&1.09\\
           &b&0.010&0.056&0.476&1.08\\
&c&          0.010&0.055&0.470&1.10\\
    &       &        & &&  \\
\hline
  \end{tabular}
\end{minipage}
\par
{\footnotesize \noindent $^{{\rm a}}$The model potential method. $^{{\rm b}}$%
The results of Chen \cite{Che94}. 
$^{{\rm c}}$ The results of Schiff {\em et al} \cite{Sch71}.} 
\end{table}

\begin{table}[h]
\caption{Oscillator strength for transitions between $^3$S and $^3$P-states
of He }
\label{tabl4}\centering
\begin{minipage}{4in}  
\vspace{1em}
  \begin{tabular}{|c|c|c|c|c|c|} \hline
     &&{  2$^3$P}  &  { 3$^3$P}  &  { 4$^3$P}& 5$^3$P \\ \hline
&&&&&\\
  1$^3$S&a&0.563&0.060&0.024&0.012\\
             &b&0.539&0.064&0.026&0.012\\
    &         c&0.539&0.064&0.026&0.013\\
    &       &        & &&  \\
2$^3$S&a&0.210&0.908&0.051&0.023\\
           &b&0.209&0.891&0.050&0.023\\
&c&          0.209&0.891&0.050&0.023\\
    &       &        & &&  \\
4$^3$S&a&0.033&0.438&1.225&0.044\\
           &b&0.032&0.436&1.215&0.044\\
&c&          0.032&0.436&1.215&0.044\\
    &       &        & &&  \\
5$^3$S&a&0.012&0.068&0.669&1.540\\
           &b&0.011&0.068&0.668&1.531\\
&c&          0.011&0.068&0.670&1.530\\
    &       &        & &&  \\
\hline
  \end{tabular}
\end{minipage}
\par
{\footnotesize \noindent $^{{\rm a}}$The model potential method. $^{{\rm b}}$
The results of Chen \cite{Che94}. 
$^{{\rm c}}$ The results of Schiff {\em et al} \cite{Sch71}.} 
\end{table}

\newpage

\newpage 

\begin{figure}[tbp]
\caption{ Laser induced couplings in metastable helium. The probe laser at
294 nm and the dressing laser at 1064 nm couple the metastable singlet state 
{2s ${}^1$S$_0$} and the initially unpopulated excited state {4s ${}^1$S$_0$}
through the ionization continuum. For strong laser intensities two-photon
ionization (IR and UV) of the metastable triplet state {2s ${}^3$S$_1$}
forms an additional background for the ion signal. }
\label{fig. 3}
\end{figure}

\begin{figure}[tbp]
\caption{ Layout of experiment, showing atomic beam, discharge, skimmer and
lasers. The helium atoms are excited in a pulsed gas discharge. The atomic
beam then interacts with the lasers in the center of a capacitor. The ions
produced there are accelerated and detected on a microsphere plate. }
\label{figlayout}
\end{figure}

\begin{figure}[tbp]
\caption{ REMPI transitions for detecting population in the metastable
states of helium. The singlet state $2s {}^1S_0$ is coupled with pulsed
radiation at 397 nm through the excited state $4p {}^1P_1$ to the ionization
continuum. The metastable triplet state $2s {}^3S_1$ is ionized with laser
light at 389 nm through the states $3p {}^3P_{0,1,2}$. }
\label{fig. 2}
\end{figure}

\begin{figure}[tbp]
\caption{Ion signal resulting from the ionization of the metastable states
after excitation of the helium atoms in a pulsed gas discharge. The
populations are probed by resonant enhanced multiphoton ionization (REMPI)
with a pulsed dye laser (bandwidth 6 GHz). The probing transitions are
strongly saturation-broadened. In the gas discharge about 90 \% of the
excited helium atoms are found in the metastable triplet state $2s {}^3S_1$,
and only 10 \% are in the singlet state $2s {}^1S_0$. }
\label{fig. 1}
\end{figure}

\begin{figure}[tbp]
\caption{Difference between ion signals with and without IR laser. The probe
laser frequency tuned far away from the two-photon resonance where LICS is
observed. The pump laser intensity is varied, while the dressing laser
intensity is kept constant at $I_S^{(0)} = 226$ MW/cm$^2$. The measured data
are normalized to the total ionization signal from singlet- and triplet
state. Good agreement is found between the experimental values and the
calculated results (dashed line). }
\label{fig. 4}
\end{figure}

\begin{figure}[tbp]
\caption{ LICS for coincident (left side) and delayed (right side) laser
pulses, when the dressing laser intensity is varied. The experimental data
are compared to results from numerical simulations (dotted lines), including
laser intensity fluctuations and integration over the spatial profile of the
pump laser. Enhancement as well as strong suppression of ionization are
observed. }
\label{fig. 5}
\end{figure}

\begin{figure}[tbp]
\caption{ Observed and calculated ionization probability of state $2s
{}^1S_0 $, as the pump laser intensity is increased. The dressing laser is
switched off. }
\label{fig. 6}
\end{figure}

\begin{figure}[tbp]
\caption{ LICS profile for weak and strong pump laser. The dressing laser
intensity is $I_d^{(0)} = 44$ MW/cm$^2$ (Fig. 8 a). The probe laser
intensity is $I_p^{(0)} = 50 $ MW/cm$^2$ (Fig. 8 b). Both laser pulses are
coincident. With increased pump laser intensity, the maximum enhancement as
well as the suppression of ionization is reduced. }
\label{fig. 7}
\end{figure}

\begin{figure}[tbp]
\caption{ An example of fitting of the experimental data using Eqn. (\ref
{ourprofile}). Squares are experimental data, dotted line is the result of
fitting, circles are the difference between experimental data and the fit ($%
\Delta t= 0$ ns, $I_p^{(0)}$= 4 MW/cm$^2$, $I_d^{(0)}$= 75 MW/cm$^2$). }
\label{fittingfig}
\end{figure}

\begin{figure}[tbp]
\caption{Width parameters vs. intensity of the dressing pulse (see Eqn. (\ref
{ourprofile}) and Table \ref{table1}): (1) is the variation of the overall
width parameter $\Gamma$, (2) is the empirically determined $\Gamma_d$, (3)
is the calculated photoionization rate $\Gamma_d$. }
\label{widthvsI}
\end{figure}

\begin{figure}[tbp]
\caption{LICS profile for two approximations: the solid line shows the
simulation results for a single pair of pulses, the dotted line shows the
profile assuming steady-state conditions ($q=0.73$, $I_p^{(0)}$= 2.5 MW/cm$%
^2 $, $I_d^{(0)}$= 38 MW/cm$^2$). }
\label{twoapprox}
\end{figure}

\begin{figure}[tbp]
\caption{Dressing pulse and delayed probe pulse ($\Delta t = -5$ ns, $%
I_p^{(0)} = 4 $MW/cm$^2$, $I_d^{(0)} = 75$ MW/cm$^2$). \protect\newline
}
\label{delayedprobe}
\end{figure}

\begin{figure}[tbp]
\caption{ Plots of $\kappa/ \kappa_0$ versus $D/\Gamma_d$ (detuning in units
of the width) for $q = 0.73$ and various values of $\Gamma_d \tau_p$. For
given $\Gamma_d$ this product is proportional to the length of the probe
laser pulse. }
\label{fig. 8}
\end{figure}

\begin{figure}[tbp]
\caption{ Probe laser frequencies at which occurs the minimum (1) and the
maximum (4) in the LICS profile of ionization $P$ versus $\Gamma_d \tau_p$.
The effect of the duration of the probe-laser pulse (with long dressing
pulse) is clearly revealed. Steady-state values (2 and 3) are shown by the
dashed lines. They approach the former for large $\Gamma_d \tau_p$. }
\label{fig. 9}
\end{figure}

\begin{figure}[tbp]
\caption{ Values of maximum (1) and minimum (2) in the LICS profile of
ionization $P$ versus $\Gamma_d \tau_p$ showing effect of pulse duration of
probe pulse. }
\label{fig. 10}
\end{figure}

\begin{figure}[tbp]
\caption{ LICS profile in various approximations. The solid line shows the
result obtained by numerically solving Eqn. (\ref{eq4}). The dashed line is
the finite-bandwidth theory, Eqn ({\ref{eq19}}). The dotted line is the
quasistationary theory, Eqn ({\ref{eq25}}). Here the dressing laser
intensity is $I_d^{(0)} = 38$ MW/cm$^2$, the probe laser intensity is $%
I_p^{(0)} = 2.5 $ MW/cm$^2$, $q=0.73$, $\tau_p=2.3$ ns, $\tau_d=5.1$ ns, the
laser pulses are coincident. }
\label{figtheory}
\end{figure}

\begin{figure}[tbp]
\caption{Photoionization cross sections for the 2$^1S$ state of He: 1 - this
work, 2 - \protect\cite{Aymar}, 3 - \protect\cite{Norcross}, 4 - 
\protect\cite{Jacobs} ( 1Mb=$10^{-18}$ cm$^2$). }
\label{fig twos photoionization}
\end{figure}

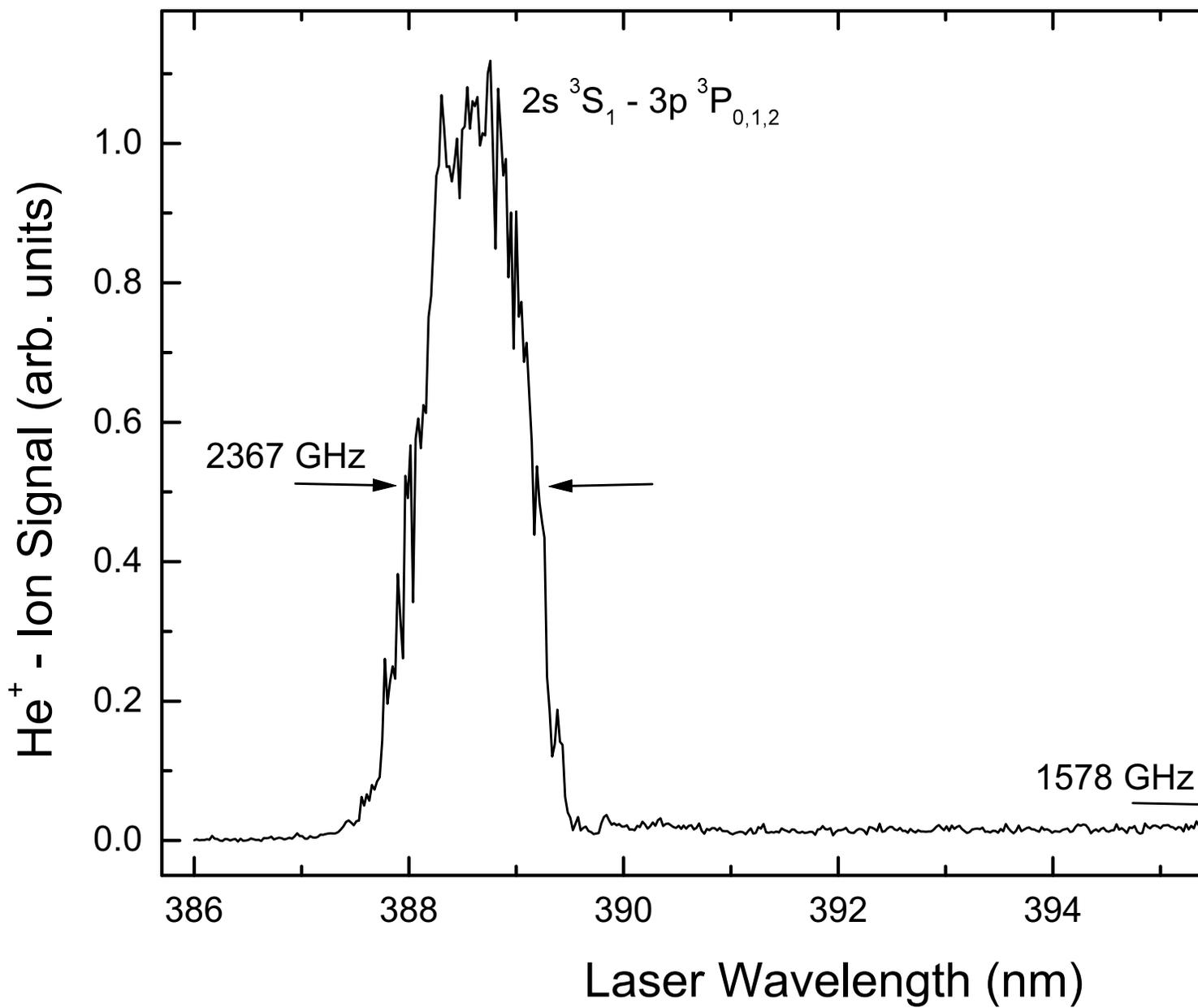
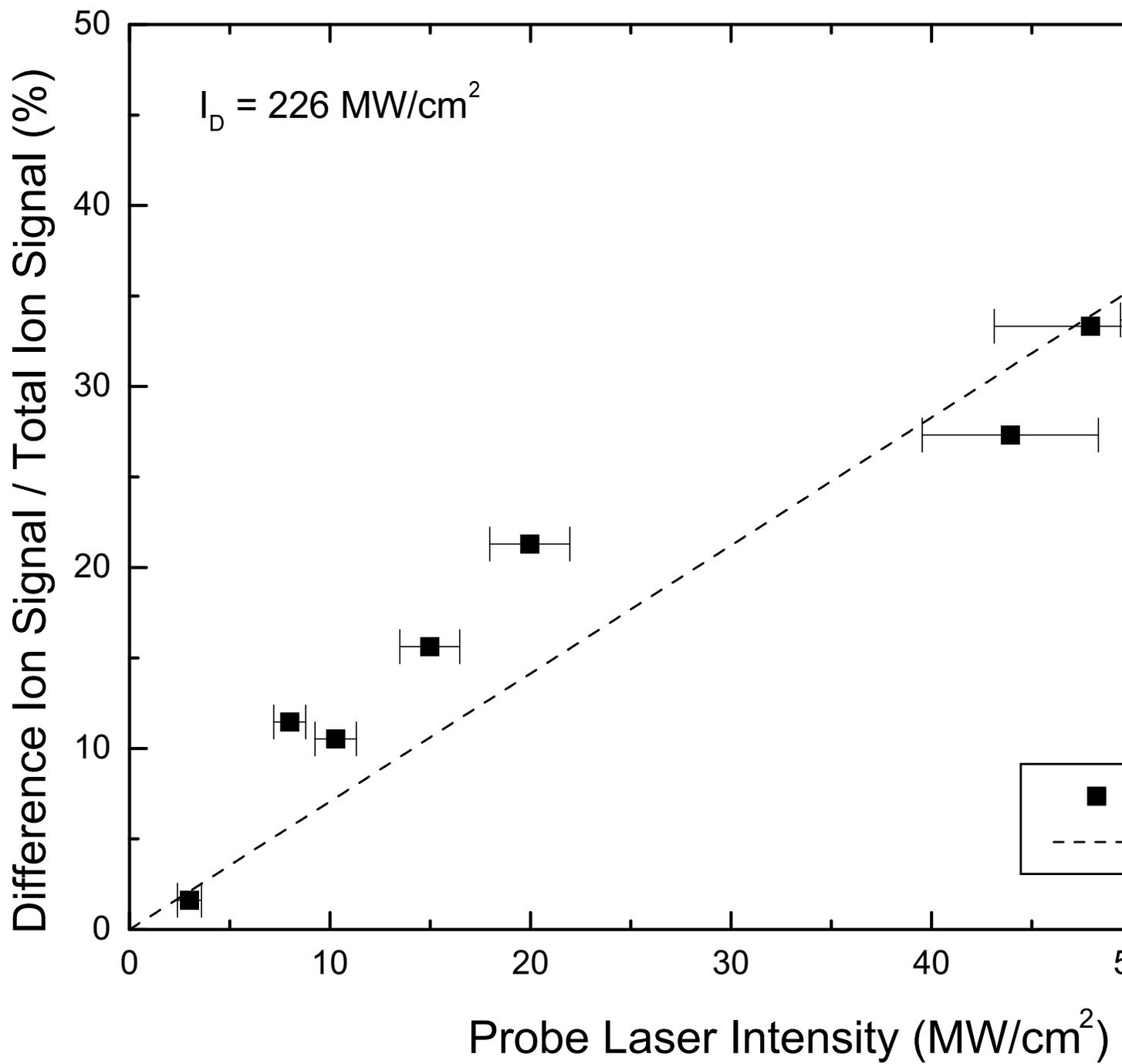
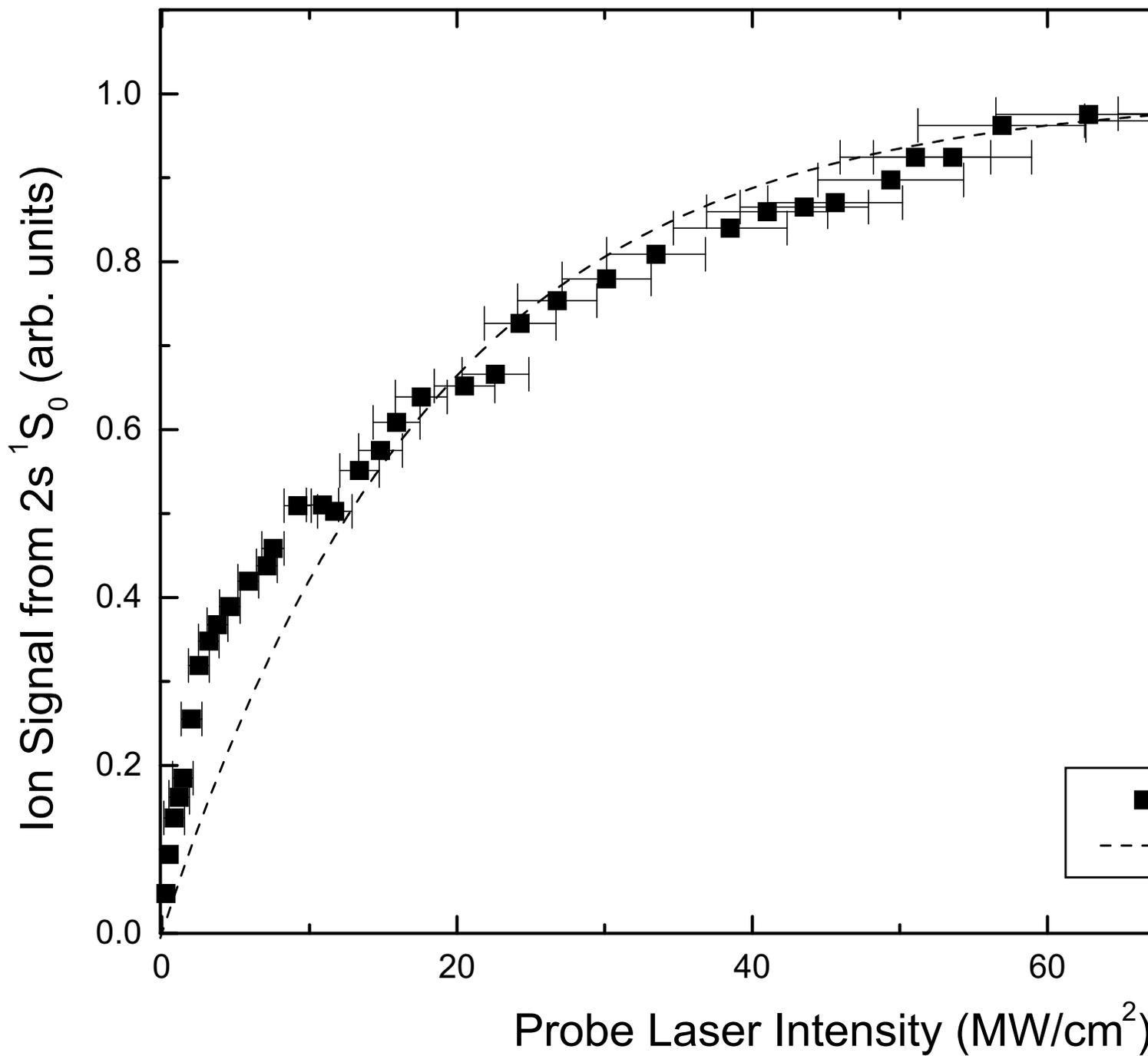
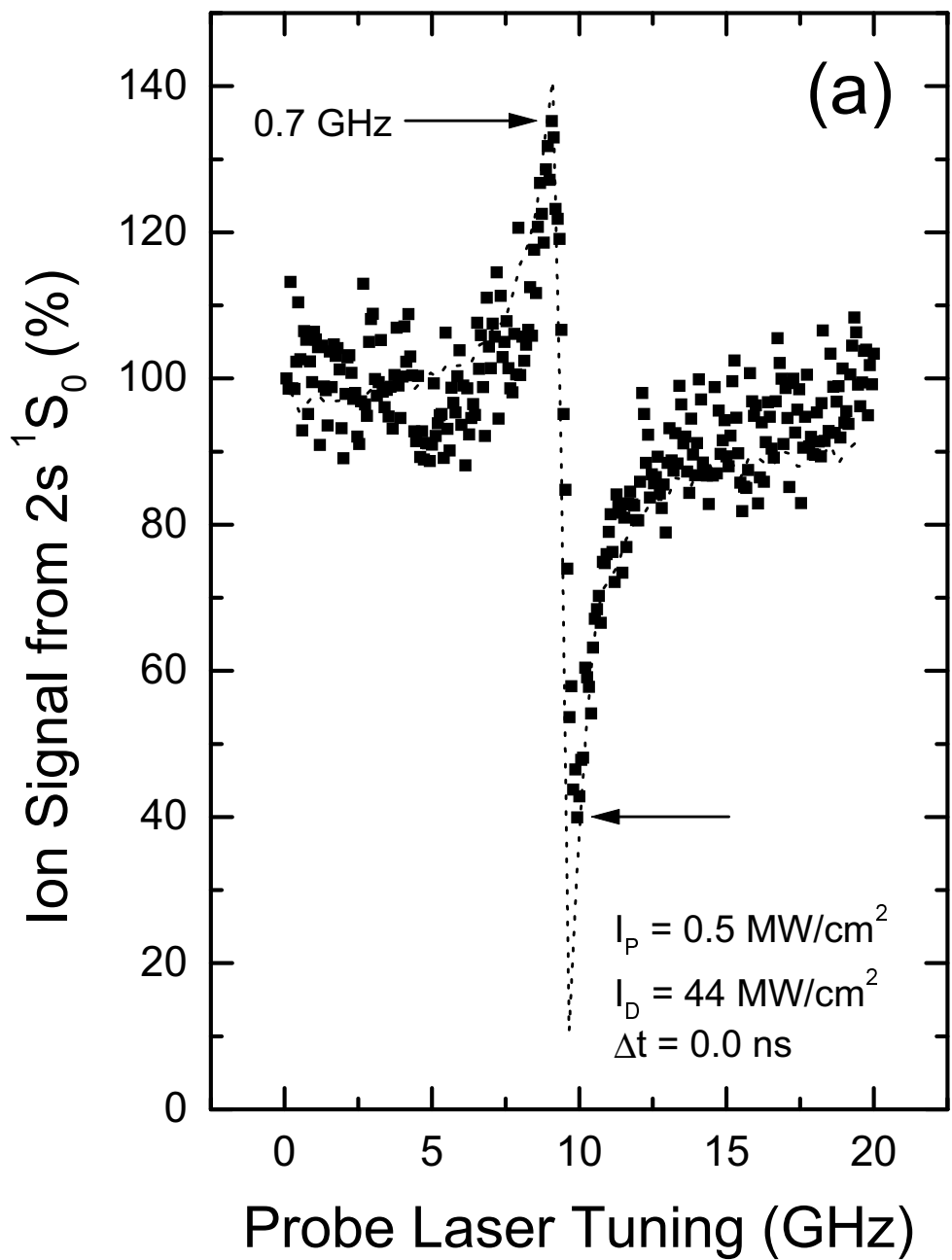
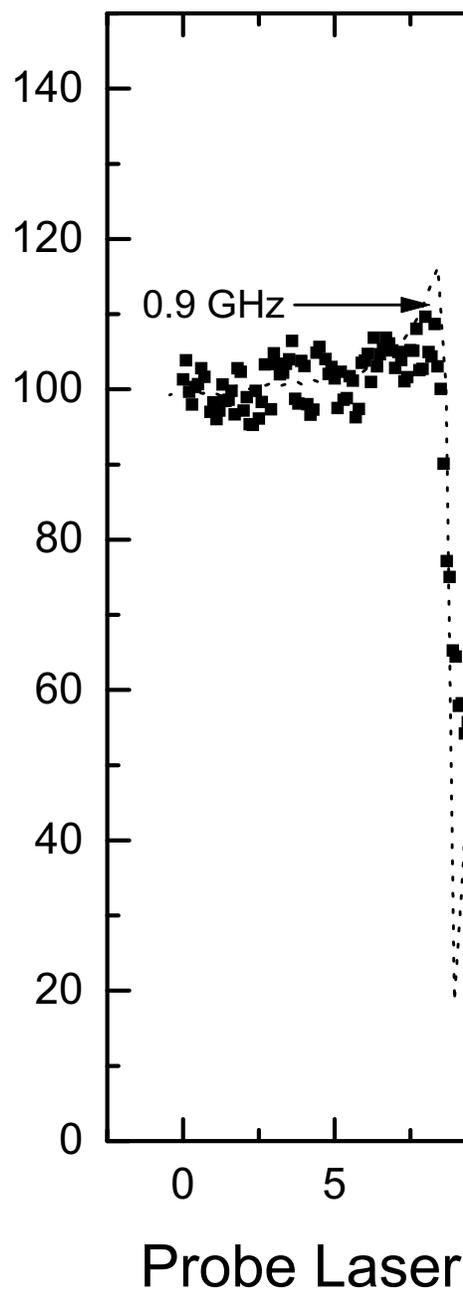
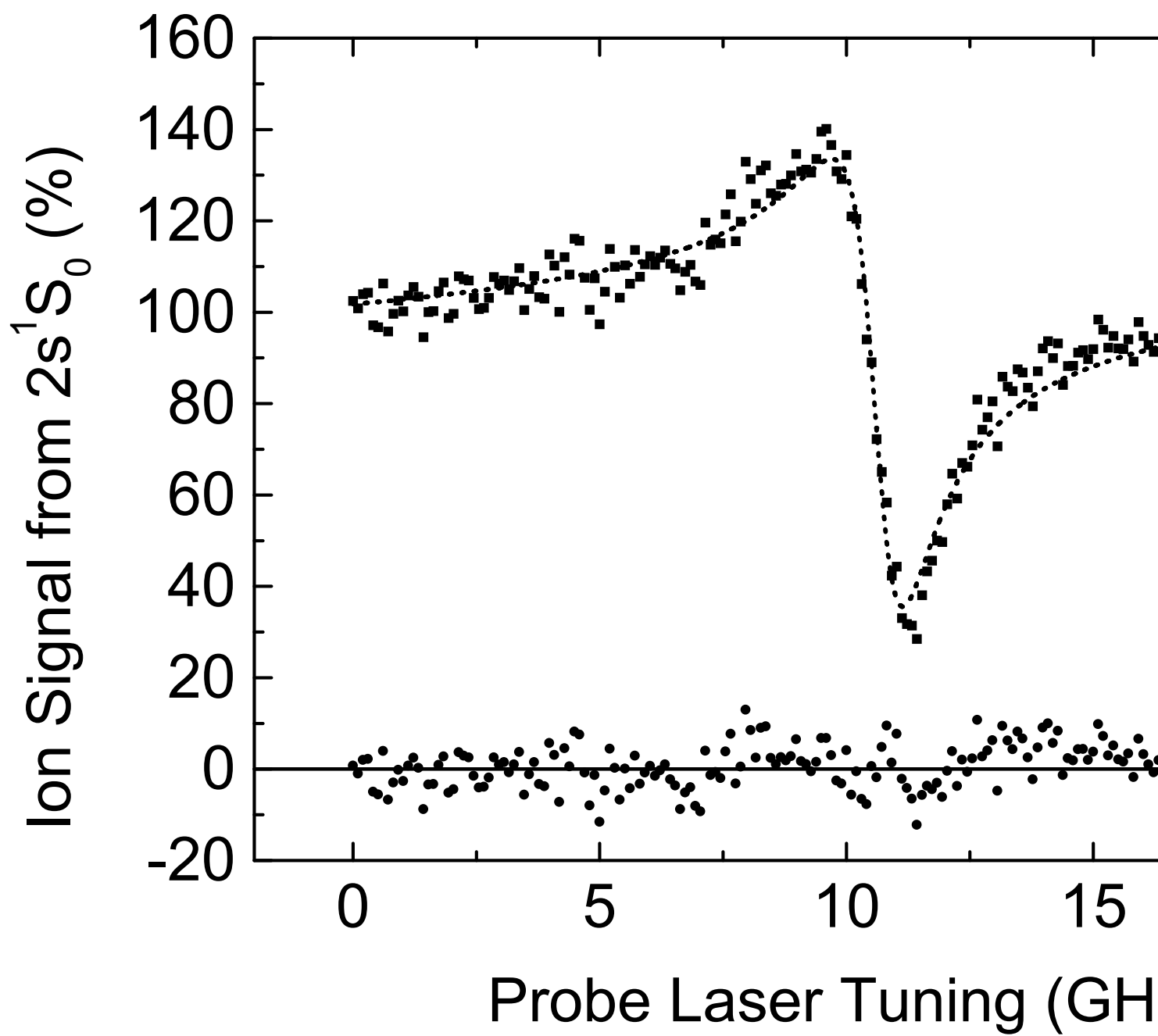
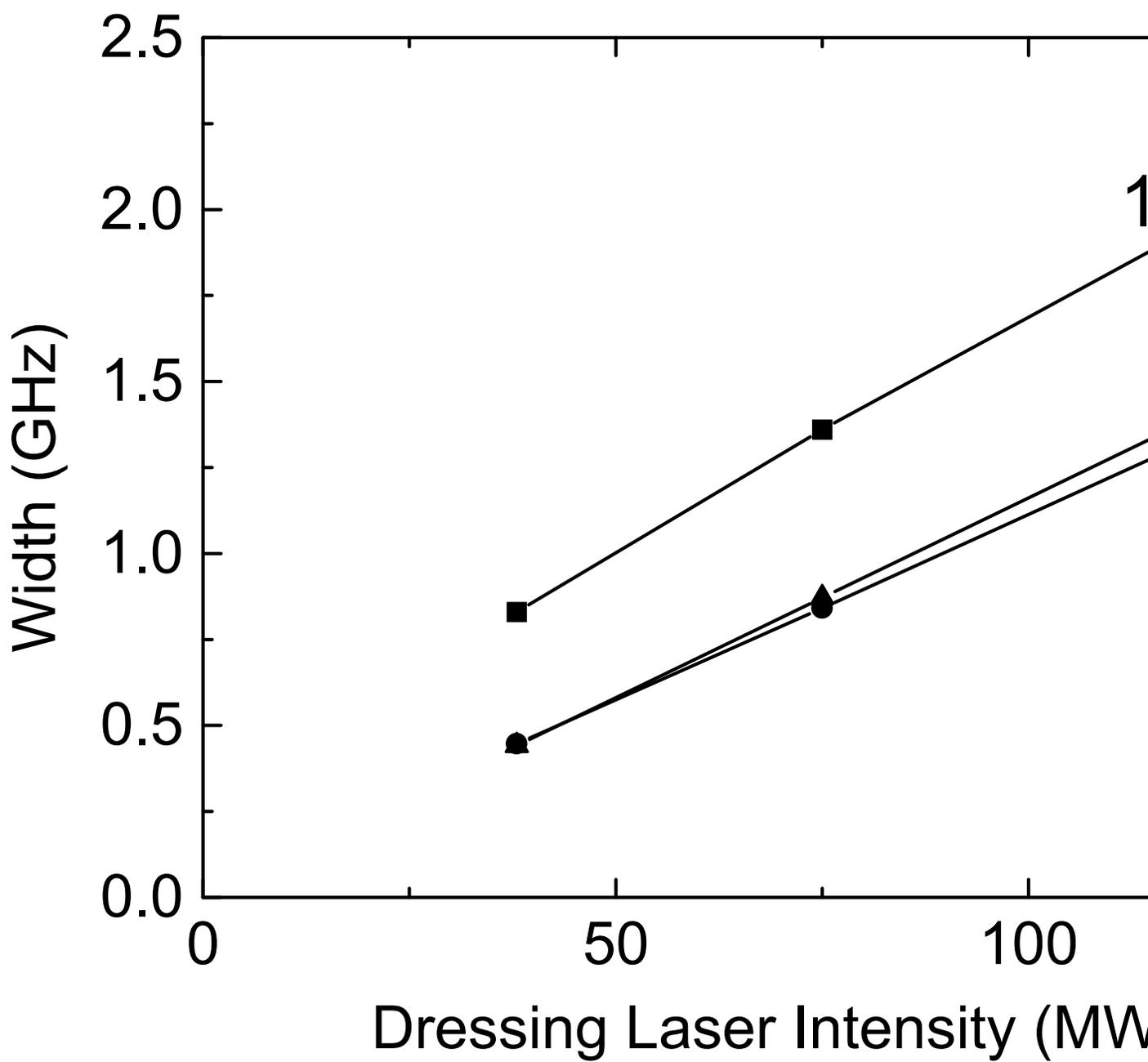
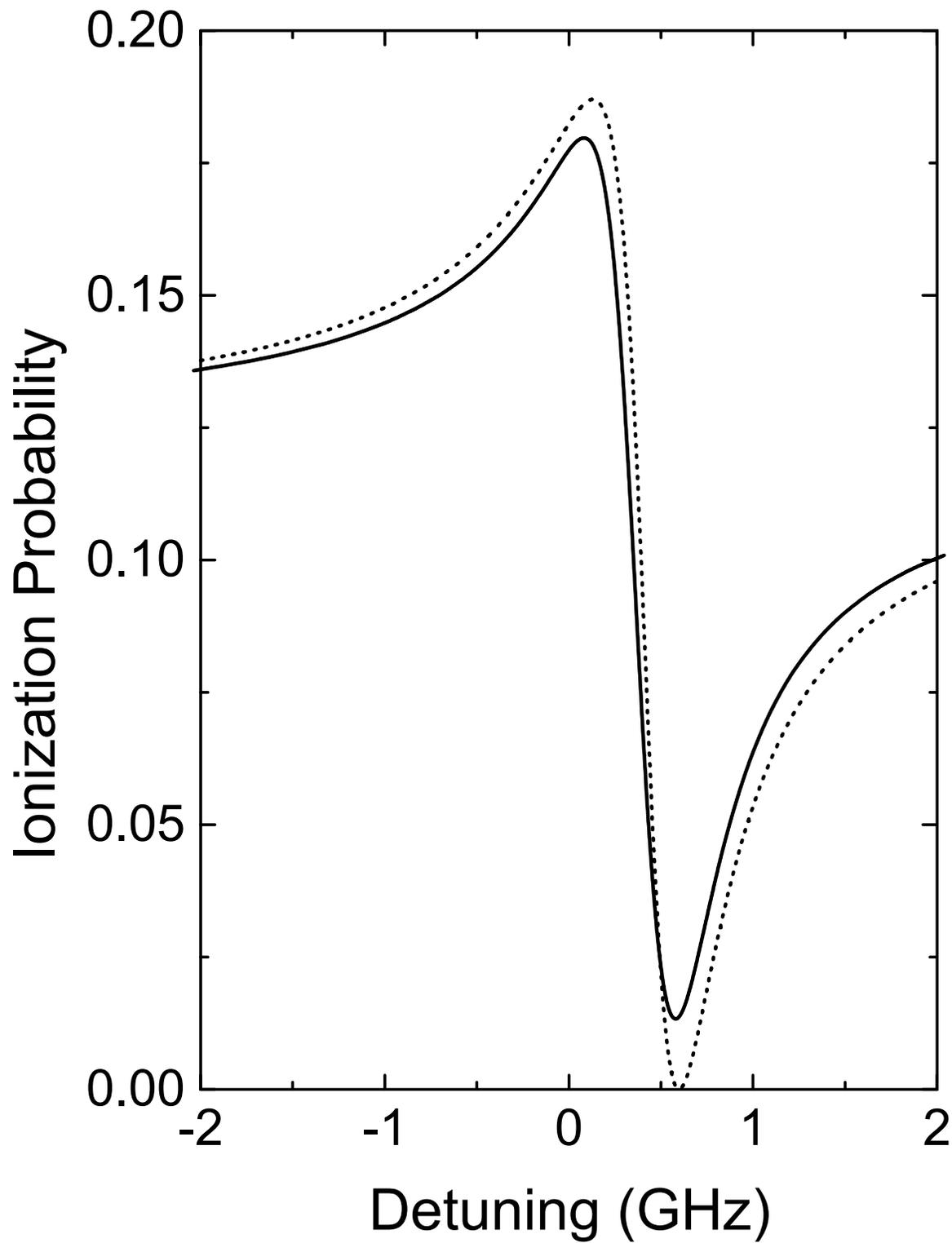
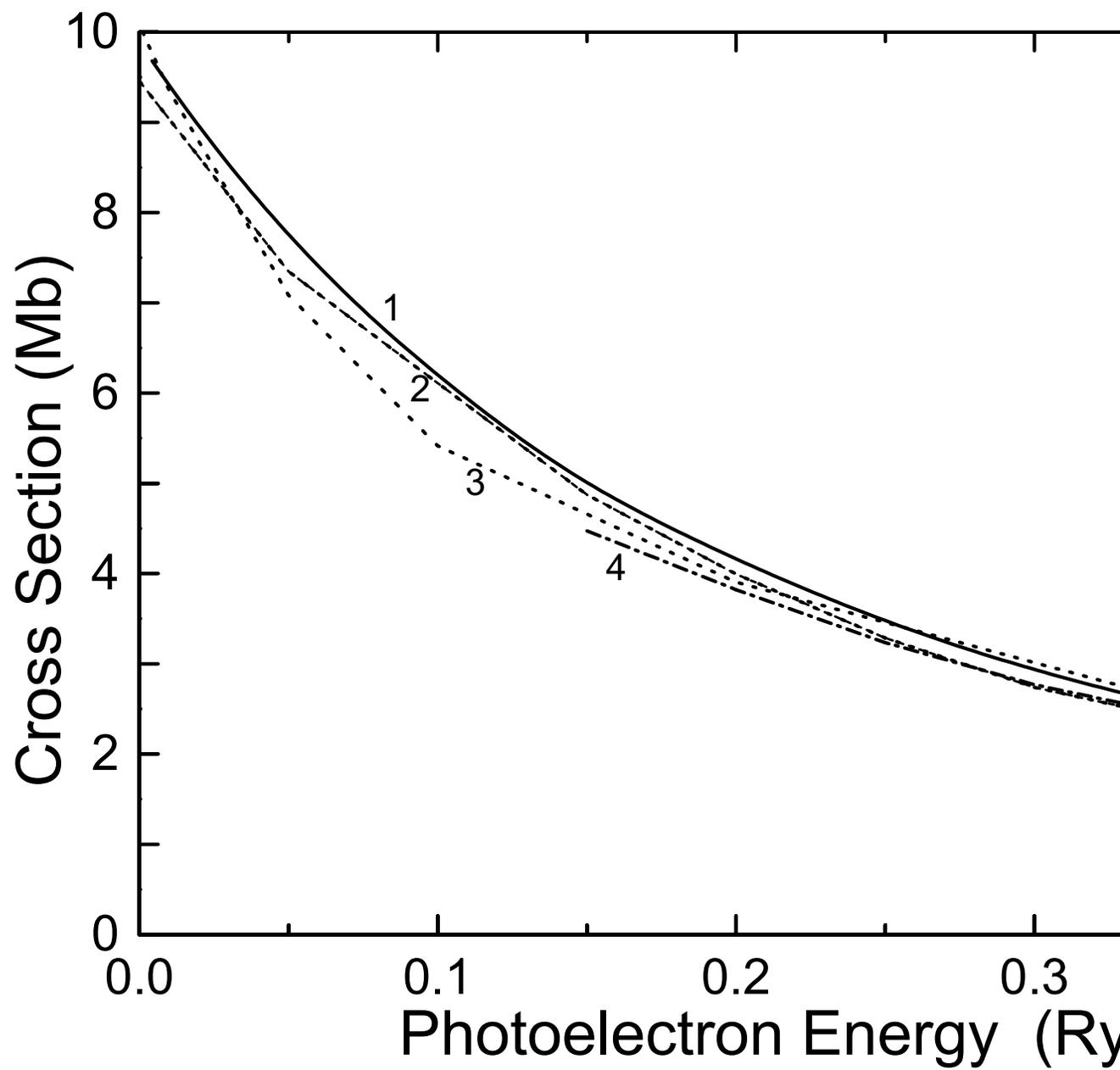
\end{document}